\documentclass[
reprint,
nofootinbib,
amsmath,amssymb,amsbsy,
prd,
onecolumn,
twocolumn,
superscriptaddress,
longbibliography,
preprintnumbers]{revtex4-1}
\usepackage[dvipsnames,usenames]{xcolor}
\usepackage{graphicx}
\usepackage{graphicx}
\usepackage{mathtools}
\usepackage[section]{placeins}
\usepackage{dcolumn}
\usepackage{bm}
\usepackage{mathrsfs,natbib}
\usepackage[normalem]{ulem}
\usepackage[title]{appendix}
\usepackage{graphicx,epsf,amssymb,amsbsy,amsfonts,amssymb,amsmath}
\usepackage[colorlinks=true]{hyperref}
\hypersetup{
    unicode=false,          
    pdftoolbar=true,        
    pdfmenubar=true,        
    pdffitwindow=false,     
    pdfstartview={FitH},    
    pdftitle={},    
    pdfauthor={},     
    pdfsubject={},   
    pdfcreator={},   
    pdfproducer={}, 
    pdfkeywords={} {} {}, 
    pdfnewwindow=true,      
    colorlinks=true,       
    linkcolor=magenta, 
    citecolor=blue,        
    filecolor=magenta,      
    urlcolor=blue           
}
\usepackage{slashed}
\usepackage{comment}

\usepackage[dvipsnames]{xcolor}
\usepackage{color,soul}
\usepackage{simplewick}

\newcommand{\be}{\begin{equation}}
\newcommand{\ee}{\end{equation}}

\newcommand{\mybar}[1]

\newcommand{\beq}{\begin{eqnarray}}
\newcommand{\eeq}{\end{eqnarray}}

\newlength{\backup}

\begin{document}
\author{Steven P.~Harris}
\email{stharr@iu.edu}
\affiliation{Department of Physics and Astronomy, Iowa State University, Ames, Iowa 50011, USA}
\affiliation{Center for the Exploration of Energy and Matter and Department of Physics,
Indiana University, Bloomington, Indiana 47405, USA}
\title{Shock-induced chiral magnetic effect}
\author{Srimoyee Sen}%
\email{srimoyee08@gmail.com}
\affiliation{Department of Physics and Astronomy, Iowa State University, Ames, Iowa 50011, USA}%

\date{June 3, 2026}

\begin{abstract}
Weak-interaction-mediated chiral imbalance generation in idealized massless electrons during core-collapse supernovae was once proposed to be the source of strong magnetic fields found in neutron stars. The effect goes by the name of chiral plasma instability (CPI). However, it was found that a finite electron mass damps out this process, inactivating the instability and preventing magnetic field growth. In this work we show that the instability can survive in the presence of abrupt density and temperature perturbation that drives the system sufficiently far out of weak equilibrium. As an example, we work with such perturbations generated by shockwaves which are common during both core collapse as well as neutron star mergers. We find that the chiral imbalance resulting from shock waves, under the right conditions of density and temperature, can sustain the chiral plasma instability despite the damping from the electron mass. Additionally, in an already magnetized medium, the chiral magnetic effect resulting from shock wave density and temperature perturbation can generate substantial ohmic heating. Our results imply that shockwaves generated in core-collapse supernovae and merging neutron stars can act as a source of strong heating in a magnetized medium as well as CPI.
\end{abstract}
\maketitle

\section{Introduction}
Weak interactions impact the evolution of dense matter in merging neutron stars \cite{Rosswog:2003rv, PhysRevLett.107.051102, PhysRevD.93.124046, Perego:2014fma, Radice:2016dwd, PhysRevD.94.123016, Baiotti:2016qnr}, protoneutron stars \cite{1986ApJ...307..178B, Pons_1999, PhysRevC.95.045807, Janka:2012wk, Reddy:1997yr}, magnetars \cite{Duncan:1992hi, Baiko:1998jq, Yakovlev:2000jp, Bedaque:2013rya}, and cold neutron stars \cite{PhysRevLett.66.2701, Yakovlev:2004yr,Yakovlev:2000jp}. They impact neutron star oscillations through bulk viscosity \cite{PhysRevC.100.035803,Harris:2024evy,Alford:2023uih,Alford:2023gxq} and produce cooling through neutrino emission. 
The simplest model for dense matter is a gas of neutrons, protons and electrons, which is charge neutral and in weak (``beta'') equilibrium
\beq
\mu_n=\mu_p+\mu_e,
\label{beta}
\eeq
where $\mu_{n/p/e}$ represents the neutron, proton and electron vector chemical potential, respectively. Note that, in weak interaction processes like beta decay or inverse beta decay in dense nuclear matter, the lightest participating particle, barring neutrinos, is the electron. Since the electron mass is much smaller than the electron Fermi energy, electrons are usually treated as massless without problem.  As is the case for massless Dirac fermions, in this case, the electrons, chirality can be treated as a good quantum number. This combined with the fact that only left chirality electrons participate in weak interactions may give the impression that electron chirality and associated parity-breaking transport effects play an important role in evolution of dense matter. Yet, for most problems of interest, this is not the case. It turns out that electron mass, despite being small, provides fast enough equilibration between electron chiralities such that any effect of parity breaking on transport arising from weak interaction is washed out in most phenomena \cite{Grabowska:2014efa}.

There can be some exceptions to this general expectation where chiral transport effects,  sourced by weak interactions can survive \cite{Yamamoto:2015gzz, Yamamoto:2015gzz, Ohnishi:2014uea, Yamamoto:2022yva, Dehman:2025jsv, Sagert:2007ug}. 
In this paper, we are specifically interested in capturing these effects: the chiral magnetic effect, to be precise \cite{Kharzeev:2013ffa, Son:2012wh, PhysRevLett.103.191601,Kaplan:2016drz, Grabowska:2014efa}. We will specifically focus on environments and parameter regimes where instantaneous equilibration of right and left chirality electrons does not apply \cite{Grabowska:2014efa, Dvornikov:2015iua, Sigl:2015xva, Kaplan:2016drz}. To quantify the transport effects, we will begin with massless electrons and then introduce the effect of electron mass through chirality flipping process.  

Clearly, in the limit of massless electrons, any weak interaction neutron decay or electron capture process will generate a chiral imbalance, which we can quantify using a chiral chemical potential, denoted as $\mu_5$. We will provide the precise definition of $\mu_5$ later. It is well known that a medium with a net chiral imbalance exhibits novel transport behavior \cite{Kharzeev:2010gd, Kharzeev:2007tn, Jiang:2015cva, Fukushima:2008xe, Son:2012wh, Manuel:2015zpa, Tuchin:2014iua, Tuchin:2025bll, Tuchin:2014iua, Mueller:2017arw, PhysRevD.97.051901, PhysRevLett.109.162001, Sen:2016jzl,Kamada:2022nyt, Avdoshkin:2014gpa}, e.g., the chiral magnetic effect (CME) \cite{Kharzeev:2013ffa, Son:2012wh, PhysRevLett.103.191601,Kaplan:2016drz, Grabowska:2014efa}: in a medium with a net chiral imbalance, there is a chiral transport coefficient which affects the electromagnetic response of the medium. In a magnetic field such a medium exhibits current along the magnetic field that is proportional to magnetic field as well as the chiral imbalance. There are two types of physical phenomena associated with this current that are relevant for dense matter, (a) the chiral plasma instability (CPI) \cite{Akamatsu:2013pjd, Kaplan:2016drz, Grabowska:2014efa,Gorbar:2021tnw} and (b) Joule/ohmic heating in a background magnetic field. CPI arises out of solving Maxwell's equations in the presence of a nonzero $\mu_5$ which leads to dynamical growth of magnetic fields. The ohmic heating response on the other hand concerns itself with background magnetic fields in the presence of a chiral imbalance. The presence of a CME current in a conducting magnetized medium found in neutron stars produces an electric field in response to the CME current. This electric field and the finite conductivity will dissipate energy through usual ohmic heating. 

While the effect of ohmic heating in dense matter due to chiral imbalance hasn't been considered widely and was proposed first in \cite{Sen:2025mzk}, the impact of CPI on the evolution of dense matter has been analyzed before \cite{Kamada:2022nyt}. 
It was conjectured in \cite{Akamatsu:2013pjd} that core-collapse supernovae could give rise to nonzero chiral imbalance by preferentially absorbing left chirality electrons in collapse. This chiral imbalance could then source chiral plasma instability effectively tying together the physics of weak interaction in core collapse with the origin of strong magnetic fields in neutron stars. However, it was shown in \cite{Grabowska:2014efa} that, assuming a constant rate for the weak interaction, electron mass can destroy chiral imbalance through helicity changing processes faster than CPI can operate. As a result, CPI in core collapse was thought to be ineffective in explaining the origin of strong magnetic fields. 

In this paper we re-assess the fate of CME in dense medium noting that while the results of \cite{Grabowska:2014efa} hold, it may not rule out CPI generating strong magnetic fields in environments where there are rapid density and temperature fluctuation. The possibility of rapid density fluctuations sustaining CPI was previously surmised in \cite{Kaplan:2016drz}.  
This type of fluctuation can drive the system far out of beta equilibrium, sustaining a chiral imbalance for long enough for the instability to generate orders of magnitude enhancement in magnetic fields. The role of nuclear matter density fluctuation in generating chiral imbalance has been previously analyzed by  Sigl \& Leite \cite{Sigl:2015xva}, who considered (turbulent) density fluctuations on length scales smaller than the neutrino mean free path.  We demonstrate that the sharp density and temperature jumps produced by shock waves can provide sufficient chiral imbalance over an extended period of time such that the instability has enough time to grow strong magnetic fields. \footnote{
Although chiral effects of neutrinos are being investigated heavily in the astrophysical context \cite{Yamamoto:2015gzz, Yamamoto:2022yva}, this work stands apart in its motivation which is to revive electron chirality driven CME that was earlier thought to be ineffective for nuclear astrophysics.} Similarly, 
we also find that these types of jumps in a magnetized medium can lead to significant Joule heating, sometimes generating thermal energy comparable to the thermal energy produced in the shockwave itself.  As shock waves can be found in neutron star mergers \cite{Dietrich:2015iva, Rosswog:2022tus, PhysRevD.111.063031, Bauswein:2010dn, Ciolfi:2017uak,Perego:2019adq} as well as core collapse supernovae \cite{Janka:2012wk, Janka:2006fh, RevModPhys.85.245,Muller:2020ard}, it is possible that weak interaction processes in core collapse supernovae are tied to strong magnetic fields through shock waves. Additionally, the Joule heating in magnetized media and CPI may also impact evolution of neutron star merger remnants due to the presence of shock waves in those environments. To answer whether this is indeed the case, one will require detailed analysis and simulations of these environments. The goal of this paper is to facilitate such future calculations by demonstrating that there are realistic conditions of temperature and density fluctuations where these effects can be substantial.  

Thus, we adopt the following approach.  We first consider uniform, non-interacting $npe$ matter that is in beta equilibrium and is at zero temperature.  We calculate the shock wave jump conditions and calculate the possible densities and temperatures of the matter as the shock passes (depending on the velocity of the fluid on either side of the shock).  We do the same thing for a couple of interacting equations of state (EoSs) in order to note the differences.  As the shock front passes through a medium, it raises the density of the downstream region by a factor of a few and also raises the temperature, potentially dramatically.  This abrupt jump in density drives the shocked fluid out of weak equilibrium.  We estimate the chiral imbalance generated in this region from the weak equilibration and consider the chirality flipping of electrons through scattering off of protons. Electron chirality flipping through scattering off of neutrons and photons is relatively smaller in the temperature regimes considered in this paper \cite{Grabowska:2014efa}. Considering an unmagnetized medium, CPI can act in this region of space right next to the shock front. As the shock front travels, this out-of-equilibrium region of space travels with it. Similarly, in a magnetized medium, this region of space which remains out of equilibrium will dissipate energy through Joule heating and as the shock front travels, this out-of equilibrium region will travel with it, heating up extended region of space in the medium.  

The organization of the paper is as follows. In section \ref{CME} we will briefly review the physics of the chiral magnetic effect, the chiral plasma instability and Joule/ohmic heating. In section \ref{sw}, we will consider shock waves and compute density and temperature jumps. We will treat this section as a place to illustrate the main ideas and make quantitative estimates that underline the importance of chiral effects. For this purpose we will restrict ourselves to two specific set of parameters upstream, denoting them as case I and case II. 
In this section, we use a non-interacting EoS for $npe$ matter as a simple illustration. The upstream parameters for the two cases are picked to match neutron star electron chemical potentials in the core. The two cases will be almost identical in terms of up stream conditions, except the upstream velocity which differentiates them. This will result in a difference in downstream thermal energy density from the shock itself.   
We will compute the chiral imbalance downstream taking into account weak interactions and chirality flipping due to the finite electron mass in subsection \ref{emass}. Additionally, we will demonstrate how the shockwave equations get modified in a constant background magnetic field in subsection \ref{bgb}.
We find that the shock-wave equations under consideration and their solutions do not change substantially in the presence of even the strongest magnetic fields we consider ($10^{18}$ G). This is relevant for the CME-induced Joule heating estimate when a shock front passes through a magnetized medium. The final two subsections \ref{cpiest} and \ref{heat} will show which of the two cases considered sustain CPI and which generate strong Joule heating. The importance of the Joule heating is quantified by comparing it to the thermal energy generated by the shock wave itself. In a subsequent section \ref{int}, we provide general results for CPI growth and Joule heating going beyond the two specific cases discussed in section \ref{sw} for non-interacting EoS. In this section, we also provide results for interacting EoS which allows us to compare the CPI growth rates and Joule heating for non-interacting and interacting EoS.  As the shockwaves themselves generate significant density and temperature increases downstream, if the matter before the shock has a density around nuclear saturation density, where the EoS is well-constrained \cite{Oertel:2016bki,Koehn:2024set,Rutherford:2024srk}, after the shock passes the matter will be at a much higher density and temperature where the EoS is less well known.  Nonetheless, it is still interesting to explore what they imply for the CPI growth and the extent of Joule heating as we do in section \ref{int}.

We work in natural units, where $\hbar=c=k_B=1$.  
\section{Chiral magnetic effect}
\label{CME}
In this section we briefly review the phenomenon of chiral magnetic effect and some of its phenomenological implications for neutron star physics. We will begin with massless electrons with a net chiral imbalance between the population of right and left chirality electrons. 
We will denote the density of right and left chirality electrons as $n_{e,R}$ and $n_{e,L}$ respectively and the corresponding chemical potentials as $\mu_{e, R}$ and $\mu_{e,L}$. In the limit of degenerate electrons, we can write 
\beq
n_{e, R/L}=\frac{\mu_{e, R/L}^3}{6\pi^2}.\label{eq:n_left_right_zerotemp}
\eeq

Of course, the above relation gets modified at finite temperature and although including finite temperature in the above relation is straightforward, for the purpose of illustration, we will work with strictly degenerate electron gas. The chiral chemical potential is defined as 
\begin{equation}
    \mu_5=\mu_{e,R}-\mu_{e,L}
\end{equation}
whereas the vector chemical potential is
\begin{equation}
    \mu_e=\frac{\mu_{e,R}+\mu_{e,L}}{2}.
\end{equation} 
The total electron density $n_e$ and the chiral/axial charge density $n_5$ are given by
\beq
n_e=n_{e,R} + n_{e,L}\nonumber\\
n_5=n_{e,R}-n_{e,L}.
\label{n}
\eeq

When $\mu_5\neq 0$, there is a chiral transport coefficient that modifies Maxwell's equations. This is known as the chiral magnetic effect (CME) and often derived using Landau level quantization for the electron.
The CME current was derived in \cite{Fukushima:2008xe, Grabowska:2014efa, Kaplan:2016drz} to be 
\beq
{\bf J}_{\text{CME}}=\xi \bf{B}
\eeq
with 
\begin{equation}
    \xi\equiv\frac{\alpha_{\text{EM}}}{\pi}\mu_5.
\end{equation}
Here $\alpha_{\text{EM}}$ is the fine structure constant. With CME current added, Maxwell's equations take the form
\beq
\bf\nabla \times \bf{B}-\frac{d\bf{E}}{dt}=\sigma \bf{E}+\xi \bf{B}.
\label{max}
\eeq 
Here, $\sigma$ stands for the electrical conductivity of the medium.

\subsection{Chiral plasma instability}
\label{cpi1}
Eq.~\ref{max} supports exponentially growing electromagnetic fields when $\mu_5\neq 0$. This is what is known as the chiral plasma instability. This instability can be easily observed by choosing the following ansatz for the gauge fields:
\beq
A_0=0,\,\, {\bf{A}}=(\hat{x} \cos(k z)-\hat{y} \sin(k z))e^{t/\tau}A_k(0)
\label{ans}
\eeq
From the ansatz it is clear that the gauge field has nonzero helicity. 
Here $A_k(0)$ corresponds to the amplitude of the helical mode with momentum $k$ at initial time $t=0$. The variable $\tau$ can be solved for by substituting Eq.~\ref{ans} in Eq.~\ref{max} and setting $\bf{E}=-\frac{d\bf{A}}{dt}$ and $\bf{B}=\bf\nabla\times \bf{A}$. The growth time-scale for the mode with wavelength $k$ is given by 
\beq
\tau=\frac{2}{\sqrt{4k(\xi-k)+\sigma^2}-\sigma}.
\eeq

Of these modes, the maximally growing one is for $k=\xi/2$. We define the corresponding growth rate as CPI growth rate 
\beq
\Gamma_{\text{CPI}}&\equiv &\left(\frac{1}{\tau}\right)\bigg|_{k=\xi/2}\nonumber\\
&=&\frac{2}{\sqrt{\xi^2+\sigma^2}-\sigma}\nonumber\\
\eeq
\begin{figure}[t!]
    \centering
    \includegraphics[width=1\linewidth]{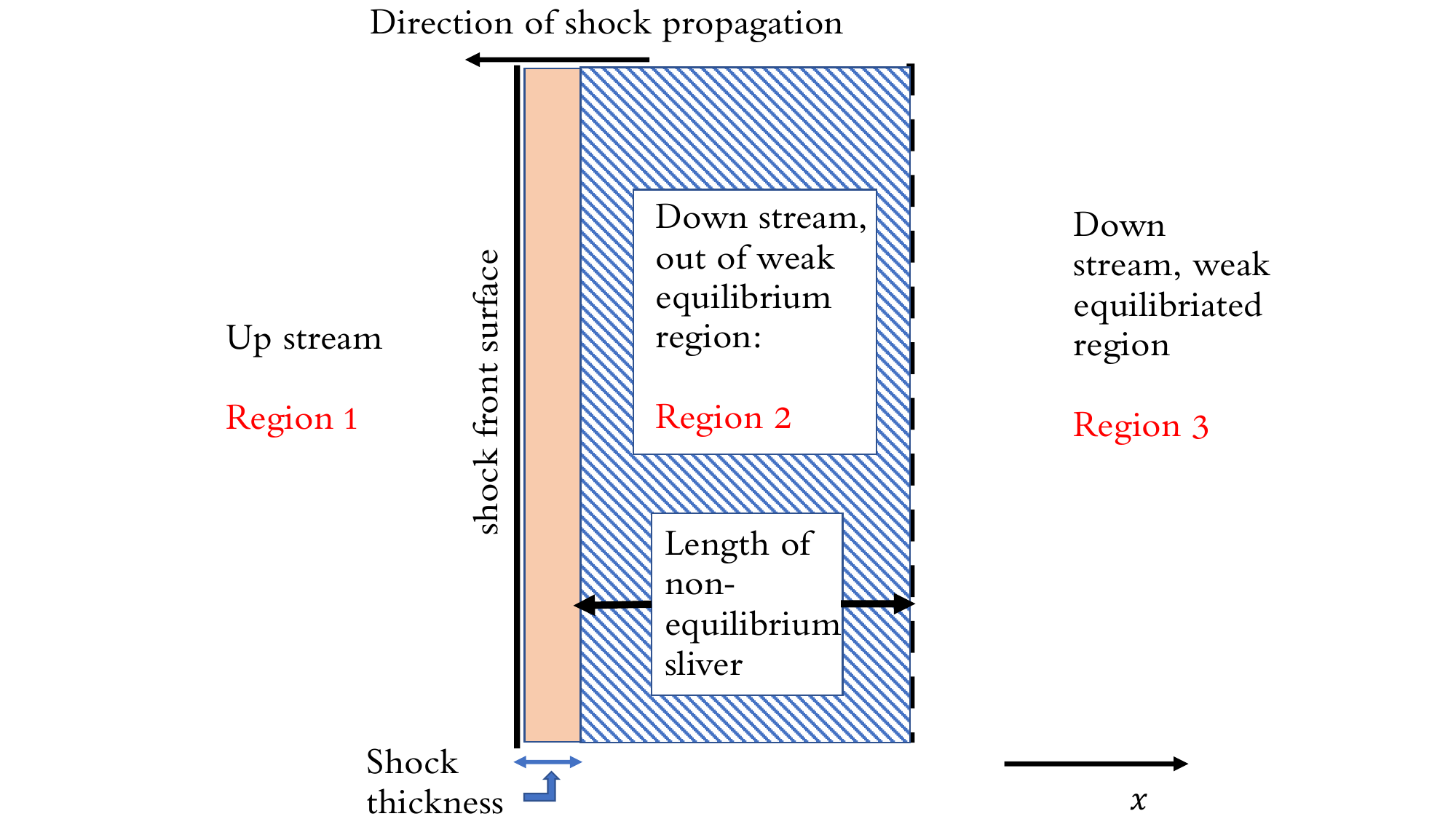}
    \caption{We illustrate the shock front in this figure which is moving to the left in the negative $x$ direction. The fluid in the up stream region (also denoted as region $1$) is yet to experience the shock whereas the two regions behind the shock front (denoted as region $2$ and region $3$) have already been shocked. The pale orange region just behind the shock surface shows the thickness of the shock. The width of this region is femtometers, set by the fluid viscosity.  Most of our analysis in this paper is in region $2$ where hydrodynamics and shock wave equations apply. However, this region is not in weak equilibrium and sustains chiral magnetic effect. Region $3$ on the other hand has achieved weak equilibrium.}
    \label{shock-pic}
\end{figure}
In dense matter relevant for neutron star physics, we see that $\xi\ll \sigma$, and the growth rate becomes
\beq
\Gamma_{\text{CPI}}&\approx &\frac{\alpha_{\text{EM}}^2}{4\pi^2}\frac{\mu_5^2}{\sigma}.
\eeq
Shocks in neutron stars mergers can heat the matter up to temperatures of several tens of MeV \cite{Perego:2019adq}, and in these conditions the protons will either be non-degenerate or semi-degenerate \cite{Hammond:2021vtv,Loffredo:2022prq}. The degeneracy temperature for protons can be estimated as
\beq
T_p\approx \frac{(3\pi^2 n_p)^{2/3}}{2M} 
\eeq
where $n_p$ here stands for the proton density. 
Thus, for $T>T_p$, protons can be considered non-degenerate and for $T<T_p$ mostly degenerate. For most of the parameter regions of interest here, we will have either $T>T_p$ or $T\sim T_p$. In this regime, the conductivity of dense $npe$ matter is given by \cite{Grabowska:2014efa}
\beq
\sigma\approx \frac{\mu_e}{4\alpha_{\text{EM}}}\frac{1}{\left(\log\left(\frac{4\mu_e^2}{m_D^2}\right)-1\right)}
\eeq
where $m_D$ is the Debye mass for the photon. The above expression can be approximated as
\beq
\sigma\approx \frac{\mu_e}{4\alpha_{\text{EM}}}\frac{1}{\log\left(\alpha_{\text{EM}}^{-1}\right)}.
\label{sig}
\eeq
Thus, we can write the CPI rate as 
\beq
\Gamma_{\text{CPI}}=\frac{\alpha_{\text{EM}}^3\mu_5^2\log\left(\alpha_{\text{EM}}^{-1}\right)}{\pi^2\mu_e}.
\label{gcpi}
\eeq
{\bf Sustained background chiral chemical potential:}
Strictly speaking, the above formula for the CPI rate $\Gamma_{\text{CPI}}$ was derived when $\mu_5$ is constant. The case of time-dependent $\mu_5$ is unexplored (in this context, at least) and deserves further study.  As a first step, in this work we will think of $\mu_5$ as a uniform background chiral chemical potential $\mu_5^b$. It is natural to ask, under what condition can a background chiral chemical potential arise. It turns out that a chiral chemical potential can arise in the presence of density fluctuation in dense $npe$ matter. This is because a rapid change in density drives matter out of weak equilibrium which preferentially absorbs or produces left chirality electrons, thus generating a chiral imbalance. However, this process by itself will lead to a growing chiral imbalance with time. We would require an accompanying mechanism that destroys chiral imbalance in order to achieve a constant background chiral chemical potential. Chirality flipping originating from the electron mass provides this mechanism. 
 For dense $npe$ matter, most of such chirality flipping happens through electrons scattering off of protons \cite{Grabowska:2014efa, Dvornikov:2015iua, Kaplan:2016drz}(see also \cite{Boyarsky:2020ani}, where processes that contribute to chirality flipping in electron-positron plasma are calculated). We can denote the chirality flip rate due to electron mass term as $\Gamma_m$. Thus, if $S_w$ is the number of left chirality electrons being produced via weak interaction per unit time, then the rate of change of chiral/axial charge can be written as
\beq
\frac{dn_5}{dt}=S_w - \Gamma_m n_5.
\label{n5rate}
\eeq
Thus a background $n_5$ can be obtained when $dn_5/dt=0$ which gives a background chiral charge density
\beq
n_5^b=\frac{S_w}{\Gamma_m}.
\label{bg}
\eeq

Using Eq.~\ref{n} one can now obtain the corresponding background chiral chemical potential, $\mu_5^b$. Of course, the net production of chirality from the beta equilibration will not continue indefinitely, and instead will cease on some timescale $t_{\text{weak}}$ when beta equilibrium is reached. $S_w$ and $t_{\text{weak}}$ are related and we will give quantitative analysis in the latter sections of this paper. Our main point here is that, the background chiral chemical potential is sustained for a time $t_{\text{weak}}$. Thus, to determine whether the CPI can give rise to strong magnetic fields, one has to determine whether $\Gamma_{\text{CPI}} t_{\text{weak}}> 1$ where $\Gamma_{\text{CPI}}$ is evaluated with $\mu_5=\mu_5^b$. If $\Gamma_{\text{CPI}} t_{\text{weak}}> 1$, the instability has sufficient time to give rise to strong fields whereas, if on the other hand $\Gamma_{\text{CPI}} t_{\text{weak}}< 1$, it does not. 
\\ 

\subsection{Joule/ohmic heating} 
\label{joule}
We will now consider Eq.~\ref{max} in the background of a strong uniform magnetic field $\bf{B}_0$ and a nonzero constant background chiral imbalance generated just as described in the previous subsection. In such a scenario, the CME current leads to charge separation and Eq.~\ref{max} supports a solution of the form 
\beq
{\bf E}_0=-\frac{\xi \,{\bf B}_0}{\sigma}. 
\label{e0}
\eeq
Thus, the CME current acts as a current source in a medium with conductivity $\sigma$ leading to energy dissipation through ohmic heating. The corresponding energy density dissipated in time $\delta t$ is
\beq
\epsilon_{\text{J}}=\frac{1}{2}\sigma {\bf E}_0^2 (\delta t)=\frac{\xi^2 {\bf B}_0^2}{2\sigma}(\delta t)
\label{th}
\eeq
where $\delta t$ is the time over which the background chiral chemical potential or the field ${\bf B}_0$ survives.
Substituting the expression for conductivity from Eq.~\ref{sig} we get
\beq
\epsilon_{\text{J}}=\frac{2}{\pi^2}\alpha_{\text{EM}}^3 \log(\alpha_{\text{EM}}^{-1})\,\frac{\mu_5^2{\bf B}_0^2}{\mu_e}(\delta t)
\label{th2}
\eeq
Note that, if the background chiral chemical potential is being sourced by weak interactions, assuming a macroscopic magnetic field that lives longer than the weak interaction time scale, $\delta t$ is set by  $t_{\text{weak}}$. In the rest of the analysis where we consider joule/ohmic heating we will work in this regime and it is supported by \cite{Kiuchi:2015sga, Palenzuela:2021gdo}. 
\noindent

\section{Sustained chiral imbalance in shock waves}
\label{sw}
\subsection{Shock wave picture}
\label{shck}
Since we are interested in chiral imbalance generated through weak interaction processes, we will be considering phenomena in which density in a region of space inside a neutron star can rise faster than weak interaction can equilibriate. While this can happen through several different mechanisms, e.g. in neutron star mergers where the density rises by several factors over length scales of a few kilometers \cite{Baiotti:2008ra,Baiotti:2016qnr}, we focus on one of the clean mechanism where the physics in question can be addressed concretely. This setting is that of shockwaves traveling through dense medium as shown in Fig.~\ref{shock-pic}. It shows a shock front, traveling in the negative $x$ direction through a dense medium as seen from the frame of the medium. In the frame of the shock, the shock surface will be at rest whereas the fluid will be moving through the shock front in the positive $x$ direction. The upstream region is denoted is region $1$ and downstream as region $2$. Assuming compression shockwave, as the fluid passes from region $1$ to region $2$, its density will see an abrupt increase. The width of this region is set by the fluid viscosity \cite{landau1987fluid}, which is a function of the mean free path of the particles involved \cite{Alford:2017rxf}.  In nuclear matter, this scale is the femtometer scale of the strong interaction \cite{Rios:2011zd}.  The mean free path of some particles, like neutrinos \cite{Espino:2023dei,Foucart:2022bth,Sawyer:1978qe}, can be much longer, but we neglect neutrinos in this analysis, deferring comment on them to the end of the conclusions section.  Thus from the point of view of the analysis here, the density jump is abrupt. 

In Fig.~\ref{shock-pic}, the colored region behind the shock front shows the shock thickness. Behind this colored region we show a region with diagonal stripes. This region corresponds to a sliver behind the shock front over which density jump has occurred and has pushed the region out of weak equilibrium. As the shock travels, this region travels with the shock. Behind the sliver, further out in the positive $x$ direction we have the region of space where weak equilibrium has been reached after the passage of the shock wave. Thus, the shock front passed through this region at an earlier time and this region has had long enough time since the shock has passed to reach weak equilibrium. The physics that we are interested in will take place in the region with the diagonal stripes. This is where we anticipate finding a background chiral chemical potential as it travels through space. If the shock front velocity is $v_{s}$, and it takes a time $t_{\text{weak}}$ for weak interaction to equilibrate, then the width of this region is given by
\beq
L_{ne}\equiv v_{s} t_{\text{weak}}
\label{eqne}
\eeq
where the subscript `ne' stands for non-equilibrium. If the shock thickness scale is denoted as $L_h$ (which we take to be set by femto-meter), then we work in a regime where $L_{ne}\gg L_h$. Note that, while the chiral effect we are interested in get realized in this thin sliver right behind the shock front, this region travels with the shock front as the front travels macroscopical distances. Thus, the chiral effects we are about to discuss should impact macroscopic distances inside neutron stars. 

\subsection{Shockwave equations with $npe$ matter}
\label{rh}
For the purpose of illustrating the physics of shockwave in a dense medium, we will begin with a dense gas of neutrons, protons and electrons. At this stage we exclude any electromagnetic field, deferring that to Sec.~\ref{bgb}.  The electrons are highly degenerate, the neutrons are mostly degenerate and the protons are taken either to be degenerate or nondegenerate, depending on the situation.  Temperature corrections to the strongly degenerate limit can be added systematically. We will begin with a description of the shock front using the Rankine-Hugoniot (RH) relations \cite{landau1987fluid}. For this part of the analysis we will not consider any electron chirality imbalance since our goal is to analyze how chirally equilibriated matter can produce chiral imbalance after undergoing shock compression. We will consider a shock front moving along the $x$-axis, in the negative $x$ direction as shown in Fig.~\ref{shock-pic}.  The two sides of the shockwave are denoted by $1$ and $2$ where $1$ is the upstream region and $2$ is the downstream region shown in diagonal stripes in Fig.~\ref{shock-pic}.  In the shock frame, let the velocity of the fluid on side $1$ be given by $v_1$ and on side $2$ by $v_2$, both will be directed along positive $x$ direction. The corresponding $4$ velocities are given by $u_r=\gamma_r(1,v_r,0,0)$ where $r$ takes values $1, 2$ for the two sides. 

Since, we are interested in the physics of region $2$, we can treat proton, neutron and electron current densities as conserved quantities across the shock front. This is to say that no composition change or only minimal composition change due to weak interaction takes place across the shock front between region $1$ and $2$.  Furthermore, we will take electrons, protons and neutrons moving with a common velocity. 
Let us denote the particle number density for neutrons, protons and electrons as $n_{n,r}, n_{p,r}$ and $n_{e,r}$ and the current density for the same as $j^{\mu}_{n,r}, j^{\mu}_{p,r}, j^{\mu}_{e,r}$, such that
$j^{\mu}_{n,r}=n_{n,r}u^{\mu}_r, j^{\mu}_{p,r}=n_{p,r}u^{\mu}_r, j^{\mu}_{e,r}=n_{e,r}u^{\mu}_r$. Here the subscript $r$ is denoting upstream and downstream regions $1$ and $2$. 

If the stress energy tensor for the fluid on the two sides is denoted as $T^{\mu\nu}_r$, then the Rankine-Hugoniot relations are given by
\beq
j^x_{p,1}=j^x_{p,2}, \,\,j^x_{n,1}=j^x_{n,2}, \,\,j^x_{e,1}=j^x_{e,2}
\label{curr}
\eeq
and 
\beq
T^{0x}_1=T^{0x}_2, T^{xx}_1=T^{xx}_2.
\label{T}
\eeq
In general we can write, 
\beq
T^{\mu\nu}_{r}= h_{r}u^{\mu}_{r}u^{\nu}_{r}- p_{r} g^{\mu\nu}
\eeq
where $h_r$ and $p_r$ are the enthalpy density and pressure on side $r$.
From here, we will used capitalized alphabet subscripts to denote the the species proton, neutron or electrons. For example, $n_{A,r}$ is the density of species $A$ on side $r$.  Note that, the stress energy tensor here corresponds to only matter fields since electromagnetic effects are assumed to be small here. The continuity equations for
$T^{0x}$ and $T^{xx}$ lead to a relation between the velocities on the two sides $v_r$. We can express the two equations in Eq.~\ref{T} as
\beq
h_1 v_1^2 \gamma_1^2 +p_1 &=& h_2 v_2^2 \gamma_2^2 +p_2\label{eq:RH_equations}\\
h_1 v_1 \gamma_1^2 &=& h_2 v_2 \gamma_2^2. 
\eeq
Solving these two leads to

\beq
v_1=\sqrt{\frac{(p_2-p_1)}{(\epsilon_2-\epsilon_1)}\frac{(\epsilon_2+p_1)}{(\epsilon_1+p_2)}}\label{v1}\\
v_2=\sqrt{\frac{(p_2-p_1)}{(\epsilon_2-\epsilon_1)}\frac{(\epsilon_1+p_2)}{(\epsilon_2+p_1)}}.
\label{v2}
\eeq
Furthermore, Eq.~\ref{curr} leads to 
\beq
\frac{v_1 \gamma_1}{v_2\gamma_2}=\frac{n_{e,2}}{n_{e,1}}=\frac{n_{n,2}}{n_{n,1}}=\frac{n_{p,2}}{n_{p,1}}
\label{fin}
 \eeq
Here we substitute Eq.~\ref{v1} and \ref{v2} on the LHS, which leads to 
\beq
\left(\frac{\epsilon_2+p_2}{\epsilon_1+p_1}\right)\left(\frac{\epsilon_2+p_1}{\epsilon_1+p_2}\right)=\left(\frac{n_{e,2}}{n_{e,1}}\right)^2
\label{ee}
\eeq
where one must remember that the density jump ratio is the same for all particles as in Eq.~\ref{fin}. We also define baryon density on the two sides with   $n_{B,r}=n_{p,r}+n_{n,r}$ such that
$n_{e,2}/n_{e,1}=n_{B,2}/n_{B,1}$.

To make further progress, we need to relate energy density and pressure to density and temperature (or chemical potential and temperature). Here for the sake of simplicity we will use non-interacting $npe$ matter (see Ch.~12 of \cite{Andersson:2019yve}). In Sec.~\ref{int}, we will include interactions in the equation of state to see their effect. For non-interacting $npe$ matter we can write $h_r=\sum_A h_{A,r}, \epsilon_r=\sum_A \epsilon_{A,r}, p_r=\sum_A p_{A,r}$. 
We write down the constitutive relations for $\epsilon_{A,r}, n_{A,r}$ and $p_{A,r}$ in terms of the chemical potential for each species and temperature. First, making no assumptions about the degree of relativity of the gas or the degeneracy of the matter, each species in a free $npe$ gas is described by
\beq
n_{A,r} &=&\frac{g}{(2\pi)^3}\int d^3p f_{A,r}(p)\nonumber\\
\epsilon_{A,r} &=&\frac{g}{(2\pi)^3}\int d^3p f_{A,r}(p)\sqrt{p^2+m_A^2}\nonumber\\
p_{A,r}&=&\frac{g}{3(2\pi)^3}\int d^3p f_{A,r}(p)\frac{p^2}{\sqrt{p^2+m_A^2}}\nonumber\\
\label{fd}
\eeq
where $f_{A,r}(p)$ is the Fermi-Dirac distribution function on side $r$ for species $A$ and $m_A$ is the mass for species $A$. We will take $m_e=0$ and $m_{p}\approx m_n\approx 940\text{ MeV}\equiv M$.

At this point, we will take specific limits to get to useful expressions. One of the regimes of interest to us is where electrons are highly degenerate, neutrons are degenerate and protons are semi-degenerate or non-degenerate. 
Thus we will often use the formula corresponding to the degenerate limit and includes finite temperature corrections up to first order in temperature measured with respect to chemical potential. For protons, we will also provide the non-degenerate formula. In the formulae presented below we will have different forms for neutron, proton gas and electron gas simply because the latter is taken to be massless ($m_e=0$) whereas the former are not. Thus we will write down separate formula to apply to the neutrons, protons and electrons. The expressions below will apply to both sides of the shockwave. So, we will also omit any reference to whether we are working in region $1$, $2$. The formula for the electron gas is 

\beq
n_{e}^{dg}&=&\frac{\mu_e^3+\pi^2 T^2\mu_e}{3\pi^2}\nonumber\\
\epsilon_e^{dg}&=&\frac{\mu_e^4+2\pi^2 T^2\mu_e^2}{4\pi^2}\nonumber\\
p_e^{dg}&=&\frac{\mu_e^4+2\pi^2T^2\mu_e^2}{12\pi^2}
\label{dgegas}
\eeq
where $\mu_e$ stands for the electron chemical potential and the superscript (dg) stands for degenerate.

For protons (semi-degenerate or degenerate) and neutrons (mostly degenerate) we can write 
\beq
n_{p/n}^{dg}&=&\frac{(\mu_{p/n}^2-M^2)^{3/2}}{3\pi^2}+\frac{T^2}{6}\frac{2\mu_{p/n}^2-M^2}{\sqrt{\mu_{p/n}^2-M^2}}\nonumber\\\nonumber\\\nonumber\\
\epsilon_{p/n}^{dg}&=&\frac{\mu_{p/n}(2\mu_{p/n}^2-M^2)\sqrt{\mu_{p/n}^2-M^2}}{8\pi^2}\nonumber\\
&+&\frac{M^4\log\left(\frac{M}{\mu_{p/n}+\sqrt{\mu_{p/n}^2-M^2}}\right)}{8\pi^2}\nonumber\\
&+&\frac{T^2(3\mu_{p/n}^3-2M^2\mu_{p/n})}{6\sqrt{\mu_{p/n}^2-M^2}}\nonumber\\\nonumber\\\nonumber\\
p_{p/n}^{dg}&=&\frac{\mu_{p/n}(2\mu_{p/n}^2-5M^2)\sqrt{\mu_{p/n}^2-M^2}}{24\pi^2}\nonumber\\
&+&\frac{M^4 \log\left(\frac{\mu_{p/n}+\sqrt{\mu_{p/n}^2-M^2}}{M}\right)}{8\pi^2}\nonumber\\
&+&\frac{T^2\mu_{p/n}\sqrt{\mu_{p/n}^2-M^2}}{6}\nonumber\\
\label{dggas}
\eeq
where $\mu_{p}$ and $\mu_n$ stand for the proton and neutron chemical potential. 

For the treatment of protons, we also provide formula for a non-degenerate gas below
\beq
n_p^{ndg}&=&2\left(\frac{MT}{2\pi}\right)^{3/2}e^{\frac{\mu_p-M}{T}}\nonumber\\
\epsilon_p^{ndg}&=& M n_p+\frac{3}{2} n_p T\nonumber\\
p_p^{ndg}&=&n_p T.
\label{ndggas}
\eeq
Here, the superscript stands for non-degenerate. 

In neutron star mergers, the matter undergoing shocks starts out cold.  Thus, we set the temperature of the upstream region to be zero, for simplicity, understanding that our results will still apply if the temperature is not strictly zero, but is small compared to the particle Fermi energies.  Entropy is produced in a shock and therefore the downstream temperature will be greater than the upstream temperature.  We use the downstream temperature as a parameter to display the set of possible downstream conditions available. The greater the upstream fluid velocity (relative to the shock), the more kinetic energy is available to be converted into downstream temperature.  

We will also assume beta equilibrium upstream (region $1$). Thus, region $1$ and region $3$ in Fig.~\ref{shock-pic} are in beta equilibrium whereas region $2$ is not (because the beta-equilibrium particle content depends on density, c.f.~Fig.~\ref{fig:xp_beq} and \cite{Harris:2024evy,Shternin:2022fii,Tsang:2023vhh}.  We will now reintroduce the subscripts denoting upstream or downstream regions $1$ and $2$, choose two specific representative values for upstream electron chemical potential, denoting them as case I and II. Since, we set upstream temperature to zero, we will not attach a subscript of $2$ for the downstream temperature in region $2$ and instead will denote it as $T$. The parameters for case I and case II are shown in Table \ref{table:caseIcaseII}.  Roughly, case I corresponds to a relatively weak shock, while case II represents a much stronger shock.
\begin{widetext}
\begin{center}

\begin{table}[]
\centering
\begin{tabular}{|c|c|c|c|c|c|c|c|c|c|}
\hline
Case: & $T_1$ {[}MeV{]} & $\mu_{e,1}$ {[}MeV{]} & $v_1$ & $T_2$ {[}MeV{]} & $v_2$ & $n_{2}/n_{1}$ & $\mu_5^b$ {[}keV{]} & $\Gamma_{\text{CPI}}t_{\text{weak}}$ & $E_{J}/E_{\text{th}}$ \\ \hline
I & 0 & 180 & 0.41 & 13 & 0.28 & 1.5 & 0.08 & $8\times 10^{-5}$ & 0.002 \\ \hline
II & 0 & 200 & 0.58 & 70 & 0.27 & 2.5 & 500 & 5.6 & 4 \\ \hline
\end{tabular}\caption{Thermodynamic and hydrodynamic parameters chosen in Cases I and II and their resultant chiral imbalance results.}
\end{table}\label{table:caseIcaseII}
\end{center}
\end{widetext}

{\bf Solving RH equations:} As mentioned, we consider cold matter getting shocked and thus set temperature to zero upstream.  Beyond this, we will also provide upstream electron chemical potential as an input. This fixes the electron density upstream. In addition, charge neutrality combined with weak equilibrium fixes the neutron and proton density upstream.  Downstream, in region $2$, neutron and electron chemical potential can be eliminated in favor of proton chemical potential and temperature by using the fact that the jump in each species equals the jump in each of the other two. We could also have eliminated the proton and electron chemical potential in favor of the neutron chemical potential or eliminated proton and neutron chemical potential in favor of the electron chemical potential. Thus, we only have to solve for two variables downstream, the temperature and the downstream proton chemical potential. Using Eq.~\ref{ee}, we can solve for downstream proton chemical potential $\mu_{p,2}$ as a function of the downstream temperature $T$. This gives a series of solutions to the RH equations. 
We in principle need one more equation to get a unique $\mu_{p,2}$ and $T$. One could use Eq.~\ref{v1} and provide an upstream velocity $v_1$ to achieve this. In this paper, we first solve Eq.~\ref{ee} to obtain $\mu_{p,2}$ as a function of $T$ and pick one of the suitable values for illustrative purposes. This solution will of course have a unique $v_1$ and $v_2$ which  we compute. Once we solve for the downstream proton chemical potential $\mu_{p,2}$ as a function of temperature we are able to evaluate the departure from weak equilibrium which we quantify with the beta imbalance
\beq
\delta\mu\equiv \mu_{n,2}-\mu_{p,2}-\mu_{e,2}.
\eeq


\noindent
{\bf Case I:} For this case we take for upstream,
$\mu_{e,1}=180$ MeV. As mentioned, the upstream temperature has been set to zero. We will assume charge neutrality which leads to a proton chemical potential of $\mu_{p,1}\approx 957$ MeV. Furthermore, beta equilibrium condition for side $1$ gives 
\beq
\mu_{p,1}+\mu_{e,1}=\mu_{n,1}
\eeq
which gives $\mu_{n,1}\approx 1137$ MeV. 

The downstream plot of beta imbalance $\delta\mu$ as a function of temperature is given by the solid lines in Fig.~\ref{fig:dmu_shock_interacting}. This plot was made using exact constitutive relations for the non-interacting gas Eq.~\ref{fd}. We choose to study two points on this plot to illustrate the main concepts. First, we choose $T\approx 13$  MeV for which $\hat{\mu}_{p,2}=\mu_{p,2}-M\approx 13.5$. Thus, the downstream protons are semi-degenerate and we use the formula of Eq.~\ref{dggas} for the protons to make estimates. Formulas for electrons and neutrons both upstream and downstream are well approximated by  Eq.~\ref{dgegas} and \ref{dggas} respectively and we use them to make future estimates.

Corresponding downstream  electron and neutron chemical potential are  
$\mu_{e,2}\approx 205$ MeV, $\mu_{n,2}\approx 1194$ MeV. These numbers lead to an estimate of 
$v_1=0.41$, and $v_2=0.28$. 
This implies that in the frame of the fluid, the shock is moving with a speed of $v_1\approx 0.41$. The density jump is found to be about a factor of $1.53$. 
Thus, for a shock front of surface $L^2$ propagating for time $\delta t$, the density is raised by factor of $1.53$ within a volume of $L^2 (v_1)(\delta t)$. Taking $L\approx 1$ km, for a shock front that has traveled for $\delta t\approx 10^{-3}$ s, the shock is able to raise the density of a region of $10 \text{ km}^3$ by a factor of $1.53$. This is also supported by reference \cite{Thierfelder:2011yi} which shows density of the merger medium rising by order one in a region of $\mathcal{O}(1\,\,\text{km})^3$ volume in a millisecond indicating the boundary of the denser region propagates with a speed that is of the order of the speed of light (albeit smaller).

\noindent
{\bf Case II:}In this case, on side  $1$, we pick $\mu_{e,1}=200$ MeV. Charge neutrality leads to $\mu_{p,1}\approx 961$ MeV. And beta equilibrium condition gives 
$\mu_{n,1}\approx 1161$. As before, we solve for $\mu_{p,2}$ and $T$ for which there is shockwave solution. Here we again focus on a specific point on this plot: taking representative values of
$T=70$ MeV and $\hat{\mu}_{p,2}=\mu_{p,2}-M=-80$ MeV. This corresponds to non-degenerate protons and thus we can use Eq.~\ref{ndggas} for the protons to make estimates. Electrons and neutrons both in region 1 and 2 are well described by Eq.~\ref{dgegas} and \ref{dggas}. The corresponding density jump is by a factor of $2.5$, $\mu_{e,2}\approx 214$ MeV. The velocities are found to be $v_1\approx 0.58$ and $v_2\approx 0.27$. 

These shock parameters are reasonable given what is currently seen in numerical simulations of neutron star mergers.  Density jumps of a factor of two or more and temperature increases of several tens of MeV are observed in merger simulations \cite{Oechslin:2006uk,Ishii:2018yjg,Perego:2019adq}.

We will now introduce the weak interaction in the problem. 
\subsection{Weak interaction and chiral imbalance}
\label{cimb}
In shocks of the type discussed above, there will be an abrupt increase of density which in general will take side $2$ away from weak equilibrium. As a result, one can consider the abrupt change in density to be instantaneous which is then followed by a longer time period during which the Urca process takes effect either producing left handed electrons or absorbing them. This will give rise to a growing chiral chemical potential lasting about as long as it takes the region behind the shock (region $2$) to reach weak equilibrium.  

In the absence of a mass term for the electrons the weak interaction will generate a growing chiral chemical potential till weak equilibrium is reached. Beyond this time, the chiral chemical potential should remain constant. When chirality flipping due to a mass term is turned on however, a background constant chiral chemical potential is maintained for the duration of the weak time scale. Beyond this time, the chirality flipping due to mass term will cause the chiral chemical potential to decay. \footnote{In principle a growing electromagnetic field helicity can feedback into the value of the chiral chemical potential when CPI is active through the chiral anomaly. For now, we will ignore this feedback since it comes into play only when the instability has been able to grow sufficiently strong EM fields. Thus, we will only try to estimate the chiral chemical potential taking into the effect of chirality flipping due to mass term.}

\subsubsection{Chirality flipping due to electron mass set to zero}
\label{cfz}
When electron mass is strictly taken to zero, chirality flipping due to the mass term vanishes.  As a result the left and right chemical potentials don't equilibrate. This limit also helps us get an estimate for how long weak interactions will proceed. We will omit the subscript for side $1$ or $2$ of the shock wave at this point since we are now only considering the dynamics of region $2$ after the shock wave has passed. Instead, we will use the subscript $i$ to denote quantities just after the shockwave has passed, but before weak equilibration, and $f$ to denote the same quantities after weak equilibration. Let's assume that the proton chemical potential, immediately after the shock wave has passed (before weak equilibration) is denoted as $\mu_{p,i}$ and after weak equilibrium is reached is given by $\mu_{p,f}$. Similarly, right electron Fermi energy after weak equilibrium is reached is denoted as $\mu_{e,R,f}$ which is the same as what it was immediately after the shockwave passed (before weak equilibration happened) $\mu_{e,R,i}$. This is of course expected since the right electrons don't participate in weak interactions. The left electron Fermi energy right after the shock wave passed (before weak equilibration) is denoted as $\mu_{e,L,i}$ and after weak equilibrium is reached is denoted as $\mu_{e,L,f}$. The neutron chemical potential follows a similar convention being denoted as, $\mu_{n,i}$, $\mu_{n,f}$ before and after weak equilibration.
Furthermore, we denote neutron and proton density as $n_{p,i}$, $n_{p,f}$ and $n_{n,i}$ and $n_{n,f}$ and right and left electron density as $n_{e,R,i}$, $n_{e,R,f}$, $n_{e,L,i}$, $n_{e,L,f}$ before and after weak equilibration. Again, note that $n_{e,R,i}=n_{e,R,f}$ here.

Our goal is to solve for the densities of each species after weak equilibrium has been reached. We will assume that the temperature remains constant during which weak equilibration takes place. As a result, solving for densities is equivalent to solving for the chemical potential for each of the species. 

To obtain the downstream quantities after weak equilibration, let's first focus on charge neutrality from which we get

\beq
n_{p,f}=n_{e,f}&=&n_{e,L,f}+n_{e,R,f}\nonumber\\
&=&n_{e,L,f}+n_{e,R,i}
\label{LRi}
\eeq
where in the last line we have used $n_{e,R,f}=n_{e,R,i}$. $n_{e,R,i}$ and the temperature downstream are known from the shockwave analysis which allows us to relate $\mu_{e,L,f}$ and $\mu_{p,f}$ using Eq.~\ref{LRi}. 
Conservation of total baryon density leads to 
\beq
 n_{p,f}+ n_{n,f}&=& n_{p,i}+n_{n,i}= n_{B,2}
\label{nb2}
\eeq
where in the last equality we note that total baryon density is conserved in the weak interaction of interest. Knowing $n_{B,2}$ from the RH equations of shockwave analysis allows us to relate $\mu_{n,f}$ to $\mu_{p,f}$. 
Finally, the beta equilibrium condition gives
\beq
\mu_{n,f}&=&\mu_{p,f}+\mu_{e,L,f},\nonumber
\label{beta2}
\eeq
yielding a third relation between $\mu_{p,f}$, $\mu_{n,f}$ and $\mu_{e,L,f}$. Thus, we have three equations Eqs.~\ref{LRi}, \ref{nb2}, \ref{beta2} to solve for $\mu_{p,f}$, $\mu_{n,f}$ and $\mu_{e,L,f}$.

We illustrate the above procedure with a few simple expressions below. For the following illustration, we are working in the limit where the temperature on the downstream region, i.e.~side $2$, although nonzero, is still small enough to keep electrons and neutrons sufficiently degenerate so that we can ignore temperature in their constitutive relations. 

Thus we can rewrite Eq.~\ref{LRi} as 

\beq
n_{p,f}&=&n_{e,f}\nonumber\\
&\approx &\frac{(\mu_{e,L,f})^3+(\mu_{e,R,f})^3}{6\pi^2}=\frac{(\mu_{e,L,f})^3+(\mu_{e,R,i})^3}{6\pi^2}\nonumber\\
\label{ee2}
\eeq
where in the last line we used $\mu_{e,R,i}=\mu_{e,R,f}$ and approximate sign is present since we are ignoring temperature contributions coming $T\ll \mu_{e,L,f}, \mu_{e,R,f}$ etc.

Substituting Eq.~\ref{ee2} in Eq.~\ref{nb2}, allows us to write 
\beq
n_{n,f}=n_B-\frac{(\mu_{e,L,f})^3+(\mu_{e,R,i})^3}{6\pi^2}.
\eeq
Since, neutrons mostly remain degenerate, we can approximate 
\beq
n_{n,f}\approx \frac{\left(\sqrt{\mu_{n,f}^2-M^2}\right)^3}{3\pi^2}
\eeq

which allows us to write 
\beq
&&\mu_{n,f}\nonumber\\
&&=\sqrt{\left(3\pi^2\left(n_B-\frac{(\mu_{e,L,f})^3+(\mu_{e,R,i})^3}{6\pi^2}\right)\right)^{2/3}+M^2}\nonumber\\
\label{munf}
\eeq
Combining the beta equilibrium condition of Eq.~\ref{beta2} with Eq.~\ref{munf} we can expresses $\mu_{p,f}$ completely in terms of $\mu_{e, L, f}$. This means we can write the proton density of Eq.~\ref{dggas} or \ref{ndggas}, whichever appropriate depending on $T>T_p$ or $T<T_p$ and substituting $\mu_{p,f}$ written as a function of $\mu_{e,L,f}$. Now, we can solve for $\mu_{e,L, f}$ using charge neutrality Eq.~\ref{ee2}.  

Once we have solved for $\mu_{p,f}, \mu_{n,f}, \mu_{e,L,f}$ and $\mu_{e,R,f}$ we can estimate how many excess left electrons are created in weak interactions 
\beq
n_L\approx\frac{1}{6\pi^2}((\mu_{e,L,f})^3-(\mu_{e,L,i})^3).
\eeq

The weak interaction process that produces or absorbs left-handed electrons in this setting is the Urca process \cite{Lattimer:1991ib,Yakovlev:2000jp}.
The net rate of the direct Urca process (the neutron decay $n\rightarrow p+e^-+\bar{\nu}_e$ rate minus the electron capture $e^-+p\rightarrow n+\nu_e$ rate) is, for free $npe$ matter in the degenerate limit, \cite{Most:2022yhe}

\beq
 S_{\textbf{Urca}} &=& \dfrac{1}{240\pi^5}G^2\left(1+3g_A^2\right)\mu_n\mu_p \mu_e\delta\mu\nonumber\\
 &&\times \left(17\pi^4 T^4+10\pi^2\delta\mu^2T^2+\delta\mu^4\right).
 \label{urc}
 \eeq
As shorthand, we will define $G\equiv G_F\cos{\theta_C}$ where $G_F$ is the Fermi constant given by $G_F=1.166\times 10^{-5} \text{GeV}^{-2}$ and $\cos\theta_C\approx 0.97$ where $\theta_C$ is the Cabbibo angle.  At the point at which we will evaluate this rate is, $\mu_e = \mu_{e,L,i}=\mu_{e,R,i}=\mu_{e,R,f}, \mu_n=\mu_{n,i}, \mu_p=\mu_{p,i}$.  
The rate of the Urca process is traditionally split into the direct Urca and modified Urca rates, where the modified Urca process includes the contributions of spectator nucleons that interact strongly with, for example, the decaying neutron \cite{Yakovlev:2000jp} (though see a recent proposal for a unified treatment of direct and modified Urca \cite{Alford:2024xfb}).  However, we neglect the modified Urca process here because for $T\gtrsim 1\text{ MeV}$, the direct Urca process dominates over modified Urca \cite{Alford:2018lhf}.  In fact, although in degenerate matter the direct Urca process is kinematically blocked if the proton fraction is too low, we neglect this kinematic ``triangle'' condition here because the downstream temperatures are $T\gtrsim 1\text{ MeV}$, for which the thermal blurring of the Fermi seas allows the particles to overcome any Boltzmann suppression and the direct Urca process proceeds uninhibited \cite{Alford:2018lhf,Alford:2021ogv}.  The net rate clearly vanishes when $\delta\mu=0$.

We note here that the rate Eq.~\ref{urc} is to be treated as an approximation. First of all, we assumed that the matter is neutrino transparent, and thus neutrinos do not participate in the initial state of any process, as their mean free path is assumed to be too long for this to be possible \cite{Alford:2018lhf,Alford:2021ogv}.  This assumption of neutrino transparency will be further discussed below.  Secondly, this expression is calculated assuming degenerate neutrons, protons, and electrons, where the phase space available to each particle grows in proportion to $T$, the thermal blurring of the Fermi surface.  As the temperature increases to several tens of MeV, the protons and then the neutrons, as previously mentioned, become nondegenerate, and so the rate does not grow as swiftly with temperature as predicted in Eq.~\ref{urc}.  On the other hand, perhaps at these high temperatures, the flattening out of the rate with temperature causes the $\delta\mu$-dependence of Eq.~\ref{urc} to dominate, leading to different equilibration dynamics than those considered in this work.  In addition, the beta equilibrium condition $\mu_n=\mu_p+\mu_e$ becomes modified as temperature increases \cite{Alford:2021ogv,Alford:2018lhf} due to neutrino trapping. While neutrino transparency is a great assumption upstream, which we assume to be at low temperature, we may expect some neutrino trapping downstream due to its higher temperature. If the neutrino mean free path downstream is much larger than the width of region $2$, $L_{ne}$, it is likely that neutrino trapping will have minimal impact on our results. On the other hand, if the two scales become comparable, we expect our results will get modified.  We discuss the possible effects of neutrino-trapping at the end of Section \ref{int}.

The reestablishment of beta equilibrium via the Urca process creates a net imbalance of of left-handed electrons compared to right-handed electrons, and the timescale over which this happens will be estimated in two ways.  The first is to assume the Urca rate is constant (i.e., the decrease in $\delta\mu$ as the equilibration proceeds is not accounted for except with an abrupt drop to zero when the constant rate completely eliminates the imbalance) and the second is to account for the dependence of the rate on $\delta\mu$, but to assume $\delta\mu \ll T$.  We use the first method in this section for illustrative purposes and the second in Sec.~\ref{int} where we give more detailed results. But the resultant timescales only differ by a factor of a few when applied to the same EoS.

The time it takes to create $n_L$ electrons through Urca process can be obtained in a straightforward way with
\beq
t_{\text{weak}}=\frac{n_L}{S_{\text{Urca}}}.\label{t_weak_1}
\eeq

We can now estimate the width of region $2$, i.e.~the downstream out of equilibrium region where weak interaction is in play, is given by
\beq
L_{ne}=t_{\text{weak}} v_1
\eeq
where we have used $v_1=-v_{\text{s}}$ (note Eq.~\ref{eqne}).

The corresponding left electron chemical potential for $t<t_{\text{weak}}$ can also be estimated as follows. Note that, we know left electron chemical potential right after the shockwave has passed, given by $\mu_{e,L,i}$ and right after weak equilibriation has been reached, $\mu_{e,L,f}$. In between these two instants in time, the rate at which they are being produced is taken to be approximately constant set by $S_{\text{Urca}}$. Thus, we can write down an approximate expression for a time dependent left electron chemical potential given by
\beq
\mu_{e,L,f}\approx \left((\mu_{e,L,i})^3+6\pi^2\left(n_L\right)\frac{t}{t_{\text{weak}}}f(t)\right)^{1/3}
\eeq
where if $t_s$ is the time at which the shock wave passed region $2$, $f(t)=\Theta(t-t_s)\Theta(t_s+t_{\text{weak}}-t)$.
This also gives an approximate expression for a time dependent chiral chemical potential given by
\beq
&&\mu_{e,L,f}-\mu_{e,R,f}
\nonumber\\
&\approx&\left((\mu_{e,L,i})^3+6\pi^2\left(n_L\right)\frac{t}{t_{\text{weak}}}f(t)\right)^{1/3}-\mu_{e,L,i}\label{chir}\nonumber\\
\eeq

\noindent
{\bf Case I:}
For case I the above procedure leads to, 
\beq
\mu_{e,L,f}&\approx &235 \,\,\text{MeV}, \nonumber\\
\mu_{e,R,f}&\approx &205\,\,\text{MeV}, \nonumber\\
\mu_{p,f}&\approx &959\,\, \text{MeV}, \nonumber\\
\mu_{n,f}&\approx& 1194 \text{MeV}\,\, \nonumber\\
\eeq

Substituting the above numbers we get $n_L$ of about $n_L^{(1)}\equiv \frac{1}{6\pi^2}(235^3-205^3)\text{MeV}^3\approx \frac{1}{6\pi^2}(163.4)^3\text{MeV}^3\approx 73673 \text{MeV}^3$ where the superscript $(1)$ on $n_L^{(1)}$ stands for case I. 

We will substitute $\mu_e\rightarrow \mu_{e,L,i}$ for the estimate $S_{\text{Urca}}$ and it estimates to 
\beq
S_{\text{Urca}}^{(1)}\approx 9\times 10^{12} \text{MeV}^3 \text{s}^{-1}
\eeq
and $t_{\text{weak}}$ (time for weak interaction to create an additional density of $n_L$ left electrons) is 
\beq
t_{\text{weak}}^{(1)}=\frac{n_L^{(1)}}{ S_{\text{Urca}}^{(1)}}\approx 8.35\times 10^{-9}\text{s}
\eeq
We can now estimate the extent or the width of the non-equilibrium region $2$, which is given by $L_{ne}^{(1)}=(v_1 t_{\text{weak}}^{(1)})\approx 1\,$m. Thus, we find that $L_{ne}^{(1)}\gg L_{h}$ if $L_h$ is set by femto-meter length scale and we are well within the regime of validity of an abrupt shock front approximation. Of course, as is evident from the context, the superscript $(1)$ stands for case I. 

\noindent
{\bf Case II:} In case II, the numbers are  
\beq
\mu_{e,L,f}&\approx& 386\,\,\text{MeV}, \nonumber\\
\mu_{e,R,f}&\approx& 214\,\,\text{MeV}, \nonumber\\
\mu_{p,f}&\approx &908\,\, \text{MeV}, \nonumber\\
\mu_{n,f}&\approx & 1294\,\, \text{MeV}.\nonumber\\
\eeq
Thus, $n_L$ is of about $n_L^{(2)}\equiv \frac{1}{6\pi^2}(386^3-214^3)\text{MeV}^3\approx 805708 \,\,\text{MeV}^3$. We next evaluate the Urca rate 
\beq
S^{(2)}_{\text{Urca}}\approx 5.6\times 10^{16} \text{MeV}^3 \text{s}^{-1}.
\eeq
The corresponding weak interaction time is
\beq
t_{\text{weak}}^{(2)}=1.7\times 10^{-11} \text{s}
\eeq
Within this time the shock front moves by 
$L_{ne}^{(2)}=v_1 \times 1.7 \times 10^{-11} \text{m} \approx 3 \times 10^{-3}$m. Again, $L_{ne}^{(2)}\gg L_h$. Of course, as is evident from the context, the superscript $(2)$ stands for case II.
\subsubsection{Introducing electron mass}
\label{emass}
In this subsubsection we will take into account the effect of chirality flipping due to electron mass. Assuming a constant rate of chirality flipping, we will no longer obtain a growing chiral chemical potential as in Eq.~\ref{chir}. We instead expect to find a sustained background chiral chemical potential over the time $t_s<t\leq t_s +t_{\text{weak}}$ if $t_s$ is the initial time at which the shock surface passes region $2$.

To obtain this background chiral chemical potential we simply note that 
\beq
\frac{dn_5}{dt}\equiv S_{\text{Urca}}-\Gamma_m n_5
\label{src}
\eeq
following Eq.~\ref{n5rate}
where $\Gamma_m$ is the rate of chirality flip \cite{Grabowska:2014efa, Sen:2025mzk} and $n_5$ is the chiral/axial charge density and $S_w$ has been replaced by the Urca rate $S_{\text{Urca}}$. Setting the LHS of Eq.~\ref{src} to zero, we find the background chiral charge density 
\beq
n_5^b=\frac{S_{\text{Urca}}}{\Gamma_m}.
\eeq
From this we can estimate the background chiral chemical potential.
In the limit when the background electron chiral chemical potential is relatively small compared to the vector chemical potential, we can write,
\beq
n_5^b&\approx &\frac{\left(\mu_e+\frac{\mu_5^b}{2}\right)^3}{6\pi^2}-\frac{\left(\mu_e-\frac{\mu_5^b}{2}\right)^3}{6\pi^2}\nonumber\\
&\approx &\frac{\mu_e^2\mu_5^b }{2\pi^2},\nonumber\\
\mu_5^b&\approx &\frac{S_{\text{Urca}}}{\Gamma_m\frac{\mu_e^2}{2\pi^2}},
\label{mu5b}
\eeq
where we used the zero-temperature relationship between $n_5$ and $\mu_5$, namely
\begin{equation}
    n_5 = \dfrac{\mu_e^2\mu_5}{2\pi^2}.
\end{equation}
We discuss finite-temperature and finite-$\mu_5$ corrections to this condition in Appendix \ref{appendix}.  For a fixed $\mu_5$, the finite-temperature expression can be up to a factor of a few times larger than the zero-temperature expression, but is typically within a factor of two.  

The chirality flip rate due to the mass term from electron-proton scattering for non-degenerate protons has the form \cite{Grabowska:2014efa, Sen:2025mzk}\\
\beq
\Gamma_m\approx\frac{\alpha_{\text{EM}}^2 m_e^2}{3\pi \mu_e}(\log\alpha_{\text{EM}}^{-1}).
\label{ndg}
\eeq
where $m_e$ is the electron mass.
For degenerate protons, $T\leq T_P$, this expression gets modified to \cite{Dvornikov:2015iua,Sen:2025mzk} 
\beq
\Gamma_m\approx\frac{\alpha_{\text{EM}}^2 m^2}{3\pi \mu_e}\log{{\alpha_{\text{EM}}^{-1}}\left(\frac{2MT}{\mu_e^2}\right).}
\label{dg}
\eeq
  In this analysis, we will mostly be in the non-degenerate or semi-degenerate region of the parameter space for protons. Thus we will be using the formula of Eq.~\ref{ndg}. We see that the background chiral chemical potential is inversely proportional the chirality flip rate. If there are mechanisms enhancing or depressing the rate, the background chemical potential will reduce or increase. In the former case, the instability and the Joule heating will get weaker and in the latter case both will get enhanced.

\noindent
{\bf Case I:}
Now we can compute the background $\mu_5^b$ for this case using Eq.~\ref{mu5b} and \ref{dg}. Substituting $M=940$ MeV, $T=13$ MeV, $\mu_e=\mu_{e,L,i}=\mu_{e,R,i}\approx 205$ MeV, $\alpha_{\text{EM}}=\frac{1}{137}$, $\mu_5^b$ for case I, denoted as $\mu_5^{(1)}$ comes out to be
\beq
\mu_5^{(1)}=\frac{S_{\text{Urca}}^{(1)}}{\Gamma_m^{(1)} \frac{\mu_e^2}{2\pi^2}}\approx 0.08 \text{keV}
\eeq
where $\Gamma_m^{(1)}$ equals $\Gamma_m$ computed for case I, i.e.~with $\mu_e=205$ MeV.\\

\noindent
{\bf Case II:} We evaluate the background chiral chemical potential for our final case, case II here. We use the formula of \ref{ndg} with $\mu_e=\mu_{e,R,i}=\mu_{e,L,i}\approx 214$ MeV.  We use $S_{\text{Urca}}^{(2)}$ to evaluate the background chiral chemical potential for this case $\mu_5^{(2)}$
\beq
\mu_5^{(2)}=\frac{S_{\text{Urca}}^{(2)}}{\Gamma_m^{(2)} \frac{\mu_e^2}{2\pi^2}}\approx .5 \text{MeV}.
\eeq
\subsection{Shockwave analysis with a background magnetic field}
\label{bgb}
At this point we have made an estimate for the background chiral chemical potential sustained in the downstream non-equilibrium region, denoted as region $2$ in Fig.~\ref{shock-pic}. Since one of our eventual goals is to estimate the Joule heating caused by this chiral chemical potential in a background magnetic field, it is natural to ask if our solutions for the RH equation need to be modified to include a magnetic field. The magnetic field in realistic dense matter environments are very complicated. However, for the purpose of extracting the essential physics, it is sufficient to consider two idealized scenarios. First is when magnetic field points in the direction of propagation of the shock-wave, in this case $x$ direction. The second is when it is perpendicular to the shock wave propagation. In both cases, there will be a direct contribution to the stress energy tensor coming from electro magnetic fields as well as an indirect contribution coming from the npe fluid's interaction with the magnetic field. The latter can be accounted for by considering Landau quantization of the fermions, especially electrons which will modify the the pressure and energy density of the npe fluid. However, we expect this modification to be small compared to the EM field independent contribution even for the strongest magnetic fields we consider here, $\sim 10^{18}$ G. 
This is because Landau quantization affects electrons most significantly and the protons less so due to its much larger
mass. In addition, the neutrons are neutral and are not Landau quantized in the conventional sense. Thus, in
quantities like the pressure of the npe fluid, which is dominated by neutron degeneracy, the effect of magnetic field will be subleading.
We of course expect neutron magnetic moment to couple to the magnetic field. However, the resulting effect on
the shock solution will be even smaller. Similarly, the energy density of the npe fluid is dominated by the mass energy density of the baryons, which again renders the magnetic field modification to energy density subleading. Since the RH equations depend on the net
  pressure and the energy density (weighted by velocity and gamma factors) of the npe fluid, and we expect the effect of magnetic field on them to be subleading compared to the magnetic field independent contribution, we expect the RH solutions originating from the latter to not get overwhelmed by the magnetic field induced modification. Since, we wish to make simple estimates, the above chain of logic naturally leads us to ignore the magnetic field modification to the npe fluids energy density and pressure. However, a more complete calculation including these effects is warranted to include subleading effects and improve the accuracy of the results. In the remainder of this section, we analyze the effect of EM field's stress energy tensor on the RH equations.

When it comes to the contribution of the EM field itself, in the former case with magnetic field directed along shock propagation, the RH equations remain unchanged and so do the estimates. For the latter case of perpendicular orientation, the RH equations do change. However, as we will see the estimates will not change by much for even the strongest fields we consider. For the latter case, we will assume a magnetic field of magnitude $B_1$ and $B_2$ pointing along the $y$ direction in region 1 and 2 in the shock frame. 

Thus, in the shock frame, there will be an electric field pointing in the $z$ direction which does not encounter a discontinuity across the shock surface and is given by \cite{Kennel:1984vf}
\beq
E_z=-v_1 B_1=-v_2 B_2.
\eeq
All other components of EM fields will be zero. 
The EM stress energy tensor is given by
\beq
T^{\text{EM}}_{\mu\nu}=F^{\mu\alpha}F^{\nu}_{\alpha}-\frac{1}{4}\eta^{\mu\nu}F_{\alpha\beta}F^{\alpha\beta}
\eeq
which in this case takes the form
\beq
T_i^{\text{EM}}=\begin{pmatrix}
\frac{(1+v_i^2)}{2} B_i^2 && v_i B_i^2 && 0 && 0\\
v_i B_i^2 && \frac{(1+v_i^2)}{2} B_i^2 && 0 && 0\\
0 && 0 && -\gamma_i^2\frac{B_i^2}{2} && 0\\
0 && 0 && 0 && \gamma_i^2\frac{B_i^2}{2}
\end{pmatrix}.\nonumber\\
\eeq
in region $i=1, 2$. 

The RH equations retain their form in terms of conservation of the the different components of the stress energy tensor. However, the stress energy tensor previously used included only matter contribution. If we relabel it as $T^{\text{m}}$, we now have to use the total stress energy tensor including the EM field contribution which is given by
\beq
\tilde{T}=T^{\text{m}}+T^{\text{EM}}.
\eeq

The corresponding Rankine-Hugoniot equations are given by 
\begin{widetext}
\beq
\left(\epsilon_1+p_1\right)v_1^2\gamma_1^2 + p_1+\frac{B_1^2(1+v_1^2)}{2}&=&\left(\epsilon_2+p_2\right)v_2^2\gamma_2^2 + p_2+\frac{B_2^2(1+v_2^2)}{2}\nonumber\\
\left(\epsilon_1+p_1\right)v_1\gamma_1^2+ v_1 B_1^2&=&\left(\epsilon_2+p_2\right)v_2\gamma_2^2+ v_2 B_2^2.
\eeq
\end{widetext}
We can solve these equations for $v_1$ and $v_2$ which can then be substituted in Eq.~\ref{fin} to find the analog of Eq.~\ref{ee} in a magnetized medium. This equation can be solved to obtain downstream conditions as before. 
Note that, the magnetic field in the fluid frame on the two sides are given by
\beq
B_i^{\text{Fluid}}=\frac{B_i}{\gamma_i}.
\eeq

We find that for the parameters of interest, the magnetic field modifications to the RH equations are negligible even for the highest magnetic fields considered in this paper, i.e.~$B_2^{\text{Fluid}}=(140 \text{MeV})^2 \approx 10^{18} \text{ G}$.

\subsection{CPI estimate from background $\mu_5$}
\label{cpiest}
Here we will analyze whether the background chiral chemical potential generated in the two cases of interest can sustain CPI for long enough to generate significant magnetic field. Note that, in this case we are working with an unmagnetized background as an initial condition which will differ from the next subsection. As mentioned before, with a background chiral chemical potential the EM field rises with exponential dependence of the form $e^{\Gamma_{\text{CPI}}t}$. This growth is sustained over the time $t_{\text{weak}}^{(a)}$, where $a$ can take values $1,2$ for the two cases we are discussing. This is the time over which weak interaction produces or absorbs left chirality electrons. Thus for sufficient growth of magnetic field, we need\footnote{This condition is derived in infinite matter with a uniform background, time-independent, $\mu_5$.  In our case, the wavelength of the growing mode is several orders of magnitude smaller than the sliver width, and thus the approximation of infinite, uniform $\mu_5$ is appropriate.} $\Gamma_{\text{CPI}}^{(a)}t_{\text{weak}}^{(a)}>1$. Thus we estimate  $\Gamma_{\text{CPI}}^{(a)}t_{\text{weak}}^{(a)}$ for the two cases at hand using the formula in Eq.~\ref{gcpi}.\\ 

\noindent
{\bf Case I:} In this case, 
\beq
\Gamma_{\text{CPI}}^{(1)}=9.62\times 10^{3} \text{s}^{-1}.
\eeq
Taking into account the corresponding weak interaction time $t_{\text{weak}}^{(1)}\approx 8.35\times 10^{-9}$ s, we find
\beq
\Gamma_{\text{CPI}}^{(1)}t_{\text{weak}}^{(1)}\approx 8\times 10^{-5}.
\eeq
Thus, in the case I, we find that the instability does not have enough time to facilitate the growth of strong magnetic fields. 

\noindent
{\bf Case II:} In this case, we find 
\beq
\Gamma_{\text{CPI}}^{(2)}=3.3\times 10^{11} \text{s}^{-1}\nonumber\\
\eeq
We already know that $t_{\text{weak}}^{(2)}=1.7 \times 10^{-11} \text{s}$ from earlier analysis. Thus,
for case II, 
\beq
\Gamma_{\text{CPI}}^{(2)}t_{\text{weak}}^{(2)}\approx 5.6
\label{3num}
\eeq
This number is sufficient for a growing EM field.  The exact number in Eq.~\ref{3num} is not important - the important piece is its order of magnitude which is in the range ${\cal{O}}(1)$ to  ${\cal{O}}(10)$ suggesting that there exist a set of parameters that can lead to significant EM field growth. If we use the exact number in Eq.~\ref{3num}, it leads to magnetic field growth of about a factor of $e^{5.6}\approx 275$. Magnetic field growth by larger factors should be achievable by varying parameters like temperature and chemical potential of different species. Similarly, a change in equation of state through inclusion of interaction or transition to quark matter might also change some of these parameters in favor of even stronger EM field growth.  
\subsection{Heating estimate}
\label{heat}
We will now make estimates for the Joule/ohmic heating corresponding to the two cases of shockwave jump. We will then compare this thermal energy density with the thermal energy density generated by the shock wave. For the latter, for free $npe$ matter we will define
\beq
\epsilon^{(a)}_{\text{th}}\equiv \sum \epsilon^{(a)}_{\text{th}, A}
\label{tha}
\eeq
where $\epsilon^{(a)}_{\text{th}, A}$ stands for the thermal energy density of species $A$ on downstream side, or side 2 for $a=1,2$ standing for case I, case II. We define $\epsilon^{(a)}_{\text{th}, A}$  as 
\beq
\epsilon^{(a)}_{\text{th}, A}=\epsilon^{(a)}_A-\epsilon^{(a)}_A(T=0).\label{eq:shock_heating_definition}
\eeq
Here $\epsilon^{(a)}_A$ is the expression for $\epsilon_{A,2}$ from Eq.~\ref{fd} where for the two cases.

\noindent
{\bf Case I:} For the Joule/ohmic heating estimate for this case, we begin with Eq.~\ref{th2} and substitute the background chiral chemical potential $\mu_5^{(1)}$ and a background magnetic field of magnitude ${\bf B}_0\approx (140 \text{MeV})^2\approx 10^{18}$ Gauss. We shall discuss the effects of lesser magnetic fields in Sec.~\ref{int}.  The ohmic heating then comes out to be
\beq
\epsilon_{\text{J}}^{(1)}&\approx &\frac{2\alpha_{\text{EM}}^3 \log\alpha^{-1}}{\pi^2}\left(\frac{(\mu_5^{(1)})^2{\bf B_0}^2}{\mu_e}\right)t_{\text{weak}}^{(1)}\nonumber\\
&\approx & (16\text{MeV})^4
\eeq
where we have used $1\, \text{s}  \approx 1.5 \times 10^{21} \text{MeV}^{-1}$.

We can now compare this energy density with the thermal energy density generated by the shockwave. This can be computed using Eq.~\ref{tha} and by substituting the parameters corresponding to side $2$ there. The thermal energy comes out to be 
\beq
\epsilon_{\text{th}}^{(1)}\approx (74 \text{MeV})^4.
\eeq
Thus we find that $\epsilon_{\text{J}}^{(1)}=0.002 \epsilon^{(1)}_{\text{th}}$ for case I, thus making the Joule heating from CME much smaller than the thermal energy generated in the shockwave\\ 

\noindent
{\bf Case II:}  The Joule heating for this case is given by 
\beq
\epsilon_{\text{J}}^{(2)} &\approx &\frac{2\alpha_{\text{EM}}^3 \log\alpha^{-1}}{\pi^2}\left(\frac{(\mu_5^{(2)})^2 \textbf{B}_0^2}{\mu_e}\right)t_{\text{weak}}^{(2)}\nonumber\\
&\approx & (256\text{MeV})^4
\eeq
and the thermal energy density generated from the shock is given by
\beq
\epsilon_{\text{th}}^{(2)}\approx (181 \text{MeV})^4.
\eeq
Thus, for case II, $\epsilon_{\text{J}}^{(2)}\approx 4 \epsilon_{\text{th}}^{(2)}$. Thus, we find that the energy dissipation from CME ohmic/Joule heating is comparable to and in fact can exceed the heating from the shock wave alone. 

Out of these two cases, Case II represents a ``stronger'' shock than case I, as both the heating and the density jump across the shock are greater.  We see that in the stronger shock, the background chiral chemical potential is higher, the CPI becomes possible, and the Joule heating is larger. In the next section, we will explore a whole range of ``cases'', first for the non-interacting $npe$ matter EoS and then for several interacting EoSs.
\section{Comparison with Interacting EoS}\label{int}
The density of neutron stars is sufficiently high that including aspects of the strong interaction is vital \cite{Oertel:2016bki}.  For example, interactions stiffen the EoS so that the EoS can support neutron stars with masses greater than $2M_{\odot}$ \cite{Oertel:2016bki,Chatziioannou:2024jsr}, matching our observations of such heavy neutron stars (see, e.g.~\cite{Fonseca:2021wxt}).  Additionally, interactions modeled by the mean-field approximation can change the particle dispersion relationships and masses, altering the kinematics of chemical reactions that proceed in medium \cite{Roberts:2016mwj,Roberts:2012um,Harris:2025isp,Oertel:2020pcg,Guo:2020tgx}. This motivates the discussion in this section which aims to assess the importance of chiral effects in shock waves including interacting EoS. There are important caveats to the results we obtain. As illustrated in the previous section, the chiral effects take place when matter is strongly out of equilibrium and is at high temperature, precisely where the EoSs are weakly constrained. Thus, it is possible that our results either overestimate or underestimate the phenomenological importance of chiral effects in a real neutron star. 

An additional point worth emphasizing is that, in the following discussion, we compare the free EoS with interacting 
$npe$ matter while fixing the upstream baryon density to a few times the nuclear saturation density. 
This choice implies that the upstream electron chemical potential in the non-interacting 
$npe$ model in the subsequent analysis is significantly smaller than in the two cases considered earlier.
Specifically, in the previous free Fermi gas analysis, the upstream electron chemical potential of the free 
$npe$ gas was fixed to values expected in realistic environments. In contrast, in the discussion below, we instead fix the upstream baryon density of the free 
$npe$ gas to realistic values.
There is no unique prescription for comparing results obtained from interacting and non-interacting EoSs, and each choice will provide complementary insights. In the following paragraphs we adopt the latter choice (of matching upstream free $npe$ baryon density to realistic values). All non-interacting $npe$ results presented below use exact formulae for density, pressure, energy density for $npe$ matter without resorting to approximations for degenerate or non-degenerate limits. 

In this section, we will model the nucleon-nucleon strong force by one-boson exchange interactions.  The corresponding equation of state will be a relativistic mean field (RMF) theory, where nucleons interact by exchanging sigma (Lorentz scalar and isoscalar), omega (Lorentz vector and isoscalar), and rho (Lorentz vector and isovector) mesons.  These mesons are approximated as frozen, assuming their mean-field values - functions of density and temperature - which contribute to the energy density and pressure of dense matter.  In this approximation, nucleons behave as free particles, but with (Dirac) effective masses ($m_*$ for the neutron and proton, taken to be equal) and effective chemical potentials $\mu_i^*$.  The electrons are treated as a free Fermi gas, as their electromagnetic interaction energy is negligible except at the very lowest densities, corresponding to the neutron star crust.    Relativistic mean field theories are reviewed in the textbook \cite{Glendenning:1997wn} and in a comprehensive catalog \cite{Dutra:2014qga}, to which we refer the interested reader.

\begin{figure}\centering
\includegraphics[width=0.45\textwidth]{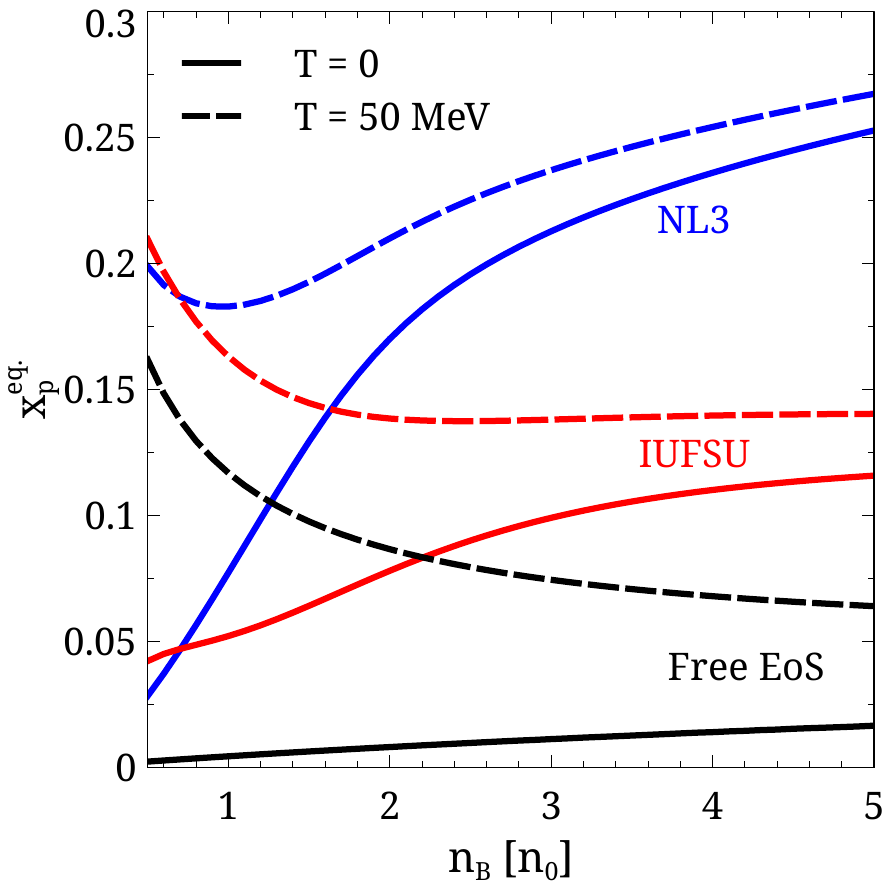}
\caption{Particle content in beta equilibrium as a function of baryon density for free $npe$ matter and for two relativistic mean field theory EoSs which include nuclear interactions.  Particle content is shown at zero temperature and at $T=50$ MeV, to show how the ``target'' proton fraction changes once the shock heating occurs.}
\label{fig:xp_beq}
\end{figure}

In this work we examine a wide variety of RMFs taken from the large collection presented in \cite{Dutra:2014qga} (see also \cite{Sun:2023xkg}).  To show the differences between the free Fermi gas EoS and realistic interacting EoSs, we plot in Fig.~\ref{fig:xp_beq} the beta-equilibrium value of the proton fraction ($x_p\equiv n_p/n_B$) predicted by two of the RMFs, both at zero temperature and at a high temperature (50 MeV) that might be reached after the matter is shocked.  The density is measured in units of nuclear saturation density, $n_0\equiv 0.16\text{ fm}^{-3}$.  Clearly, interactions favor increased proton content at a given density compared to $npe$ matter without interactions.  As is well known, the proton fraction is controlled by the density-dependence of the nuclear symmetry energy, which is the difference between pure neutron matter and the symmetric nuclear matter ($x_p=1/2$) that nearly appears in atomic nuclei - see Ch.~12 of \cite{Andersson:2019yve} or, e.g., \cite{Lattimer:2023rpe}.  The two RMFs that we show here, NL3 \cite{Lalazissis:1996rd} and IUFSU \cite{Fattoyev:2010mx}, bookend reasonable behavior of the symmetry energy.  Finally, we see from Fig.~\ref{fig:xp_beq} that the chemical composition of beta-equilibrated matter is different in hot and cold nuclear matter - the proton fraction is higher at higher temperature, and increased temperature effects the low density part of the EoS first.

\begin{figure}\centering
\includegraphics[width=0.45\textwidth]{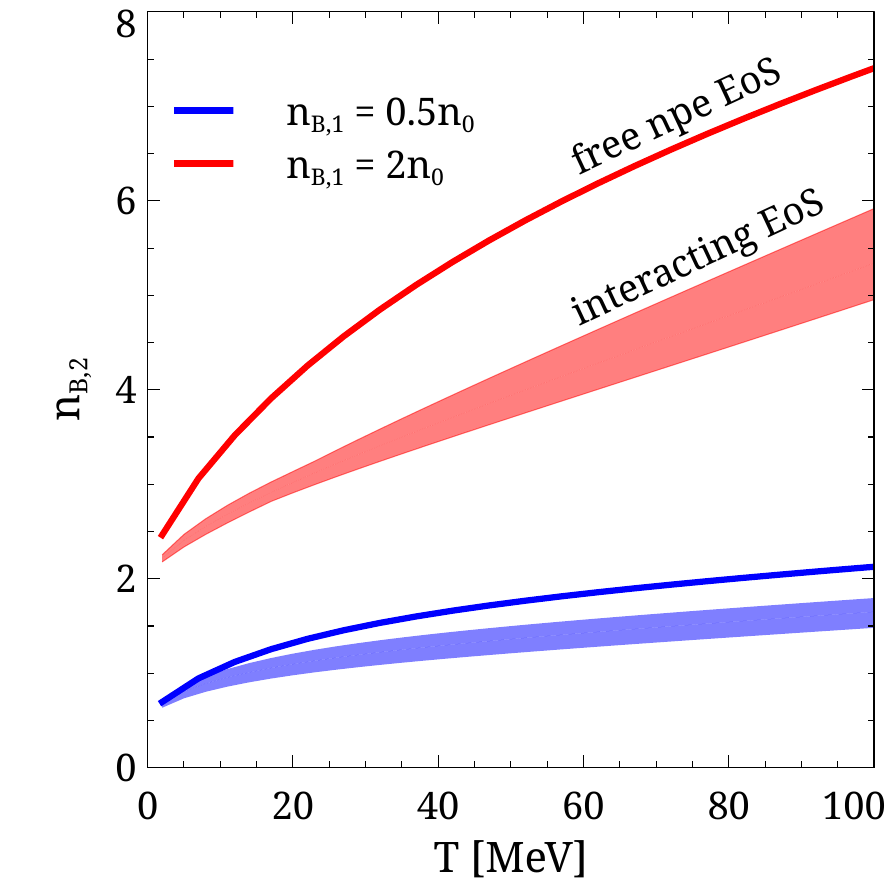}
\caption{Downstream baryon density as a function of the temperature jump.  The thin lines correspond to the predictions of a free $npe$ matter EoS, while the bands span the answers predicted by various interacting RMF EoSs.  The densities referred to in the plot legend are the baryon densities of the upstream matter.}
\label{fig:n2_over_n1_interacting}
\end{figure}

With the RMF EoS, we again specify an upstream density, temperature (zero), and impose beta equilibrium, and then solve the RH equations for different choices of downstream temperature.  In Fig.~\ref{fig:n2_over_n1_interacting}, we show the density jump across the shock as a function of the downstream temperature.  The results for a free $npe$ matter are shown in thin lines and the bands correspond to the results given by the interacting RMF EoSs.  For a given density jump, the interacting EoSs predict larger shock heating, and the shock heating increases with the size of the density jump.  

\begin{figure}\centering
\includegraphics[width=0.45\textwidth]{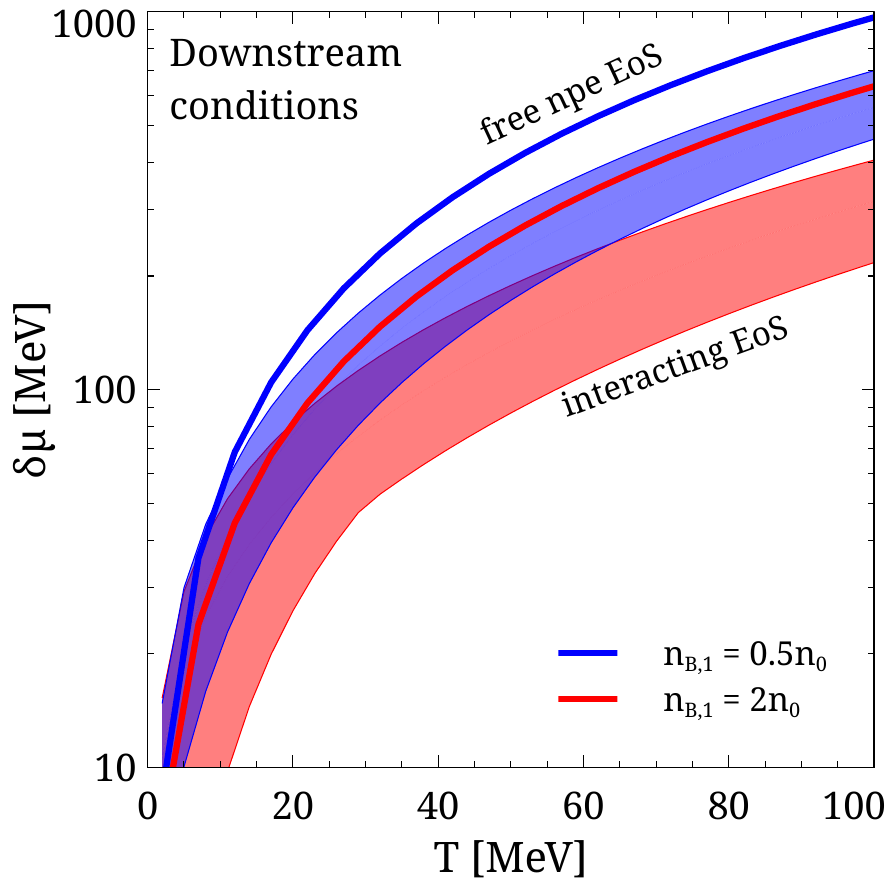}
\caption{Deviation from beta equilibrium caused by the shock, as a function of the temperature increase due to the shock. The same conventions apply as in Fig.~\ref{fig:n2_over_n1_interacting}.}
\label{fig:dmu_shock_interacting}
\end{figure}

In Fig.~\ref{fig:dmu_shock_interacting}, we show how far the fluid is pushed out of beta equilibrium due to the shock.  If the beta equilibrium value of the proton fraction was flat with respect to density as well as temperature (c.f.~Fig.~\ref{fig:xp_beq}), then $\delta\mu=0$.  We see that for a given temperature jump across the shock, the free EoS is pushed further out of beta equilibrium.  This is counterintuitive, because the beta-equilibrium proton fraction for the interacting EoSs is much more density-dependent than for the free EoS, but it is also true that for a given temperature jump, the free EoS has a higher density jump.  In any case, a shock that heats matter by several tens of MeV will push it out of beta equilibrium by 100 or more MeV.  Of course, we have assumed that weak interactions do not take place within the shock surface, as they are too slow.  

As a complement to the two cases presented in Sec.~\ref{sw}, we will provide general formulas for the background $\mu_5$, the CPI exponent, and the amount of Joule heating.  To make the formulas simpler, we will make some approximations, discussed below.  First, we derive the second method of estimating the beta-equilibration timescale (the first method was shown in Eq.~\ref{t_weak_1}).

The evolution of the left-handed electron density in a fluid element follows the equation 
\begin{align}
    \dfrac{\mathop{dn_{e,L}}}{\mathop{dt}}&=S^{\text{net}}_{\textbf{Urca}}\approx \lambda\delta\mu = \lambda\dfrac{\partial\delta\mu}{\partial n_{e,L}}\bigg\vert_{n_B}\delta n_{e,L},\label{eq:neL_evolution_equation}\\
    \lambda &= \dfrac{17}{240\pi}G^2\left(1+3g_A^2\right)\mu_n^*\mu_p^*\mu_e T^4. 
\end{align}
In the interacting case, the chemical potentials of the nucleons in the Urca rate become the effective chemical potentials $\mu_n^*$ and $\mu_p^*$, and we have approximated the net Urca rate as $\lambda\delta\mu$, known as the \textit{subthermal} limit $\delta\mu\ll2\pi  T$ \cite{Haensel:2002qw}.  We take the subthermal limit purely to generate relatively simple expressions in the following, but we note that the subthermal limit is valid\footnote{However, as we noted in Sec.~\ref{cimb}, at high temperatures the temperature-dependence of the beta equilibration rate Eq.~\ref{urc} is likely reduced due to non-degeneracy, and therefore the equilibration dynamics may in fact be suprathermal \cite{Haensel:2002qw}.  We defer such an exploration to future work.} because the temperature in the shocked matter is so large.  The derivative in Eq.~\ref{eq:neL_evolution_equation} is a susceptibility of the EoS \cite{Harris:2024evy} and is negative, and thus this equation indicates that the left-handed electron population relaxes to its beta-equilibrium value at a timescale 
\begin{equation}
    t_{\text{weak}} = \left\vert  \lambda\dfrac{\partial\delta\mu}{\partial n_{e,L}}\bigg\vert_{n_B} \right\vert^{-1}.
\end{equation}
For a degenerate free Fermi gas of $npe$ matter where $k_{Fn}, k_{Fp}, k_{Fe}$ stand for the Fermi momentum of neutron, proton and electrons,
\begin{align}
    \dfrac{\partial\delta\mu}{\partial n_{e,L}}\bigg\vert_{n_B}&=-\pi^2\left(\dfrac{1}{\mu_nk_{Fn}}+\dfrac{1}{\mu_pk_{Fp}}+\dfrac{1}{\mu_ek_{Fe}}\right)\label{eq:B_susc_analytic}\\
    &\approx -\dfrac{\pi^2}{\mu_e^2}.\nonumber
\end{align}
This quantity is clearly related to the inverse of the density of states at the Fermi surface of a degenerate Fermi gas \cite{coleman2015introduction,1970ApL.....5...33J}.  One could calculate this susceptibility consistently with an EoS \cite{Harris:2024evy}, or near saturation density with knowledge of the nuclear symmetry energy \cite{Harris:2025ncu}, but in-line with our goal of simple estimates, we stick with Eq.~\ref{eq:B_susc_analytic}.  The timescale for left-handed electron beta equilibration is
\begin{equation}
    t_{\text{weak}} \approx \dfrac{240}{17\pi}\dfrac{\mu_e}{G^2\left(1+3g_A^2\right)\mu_n^*\mu_p^*T^4}.\label{eq:t_weak_interacting}
\end{equation}
This timescale is independent of $\delta\mu$, as the subthermal limit was taken.

\begin{figure}\centering
\includegraphics[width=0.45\textwidth]{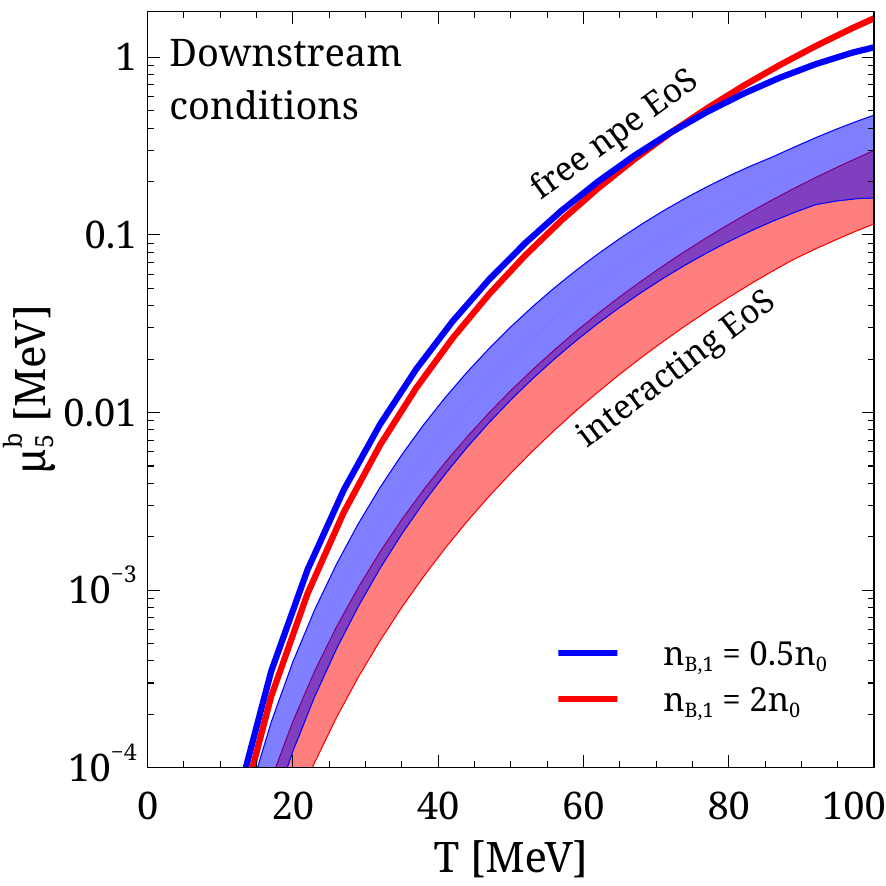}
\caption{The background chiral chemical potential $\mu_5$ sustained due to a balance between the Urca production of a chiral imbalance and the chirality-flipping decay of the imbalance due to the finite mass of the electron.  The same conventions apply as in Fig.~\ref{fig:n2_over_n1_interacting}.}
\label{fig:mu5_shock_interacting}
\end{figure}

Now, from Eq.~\ref{mu5b}, with Eqs.~\ref{ndg} and \ref{eq:neL_evolution_equation}, we can derive a formula for the background $\mu_5$
\begin{equation}
    \mu_5^b \approx \dfrac{17\pi^2}{40}\dfrac{G^2\left(1+3g_A^2\right)\mu_n^*\mu_p^*\delta\mu T^4}{m_e^2\alpha_{\text{EM}}^2\log{\alpha_{\text{EM}}^{-1}}}.\label{eq:mu5_interacting}
\end{equation}
Evidently, the background chiral chemical potential is largest for the most dramatic of shocks - shocks that push the matter maximally out of beta equilibrium and that generate the most heating of the matter.  Also, if the electron mass were larger, the chirality flipping would be faster and $\mu_5^b$ would be smaller.  

In Fig.~\ref{fig:mu5_shock_interacting}, we plot the background chiral chemical potential that results from the competition between Urca and electron chirality-flipping scattering.  We see that the interacting EoS predicts lower $\mu_5^b$ values than does free $npe$ EoS.  However, stronger shocks produce a higher background $\mu_5^b$ and particularly strong shocks are able to produce chiral chemical potentials of over 100 keV.

\begin{figure}\centering
\includegraphics[width=0.45\textwidth]{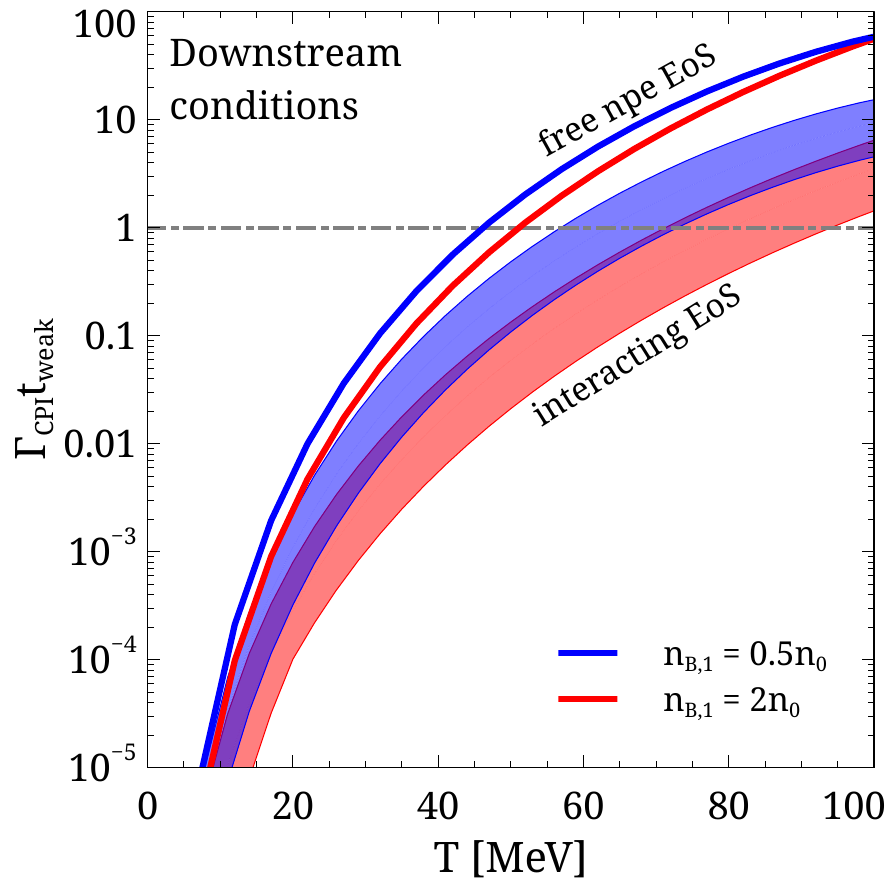}
\caption{The rate of the chiral plasma instability times the timescale over which it is sustained (the beta equilibration timescale).  Values larger than 1 indicate the possibility of a chiral plasma instability.  The same conventions apply as in Fig.~\ref{fig:n2_over_n1_interacting}.}
\label{fig:CPI_exponent}
\end{figure}

We calculate $\Gamma_{\text{CPI}}t_{\text{weak}}$ using Eqs.~\ref{gcpi}, \ref{eq:t_weak_interacting}, and \ref{eq:mu5_interacting}
\begin{equation}
    \Gamma_{\text{CPI}}t_{\text{weak}}\approx \dfrac{51\pi}{20}\dfrac{G^2\left(1+3g_A^2\right)\mu_n^*\mu_p^*\delta\mu^2T^4}{\alpha_{\text{EM}}\log{\alpha_{\text{EM}}^{-1}}m_e^4}.\label{eq:CPI_t_interacting}
\end{equation}
In Fig.~\ref{fig:CPI_exponent}, we plot the quantity in the exponent of the chiral plasma instability calculation: the CPI rate times the duration that it is active, namely, the beta-equilibration timescale.  If this quantity is greater than one, an instability develops.  We see that sufficiently strong shocks are able to generate the conditions for the CPI.  In particular, the interacting EoSs indicate that low-density matter that is shocked up to temperatures of over 50 MeV can lead to a high enough $\mu_5^b$ sustained over a long enough time for the CPI to develop.

\begin{figure*}\centering
\includegraphics[width=0.45\textwidth]{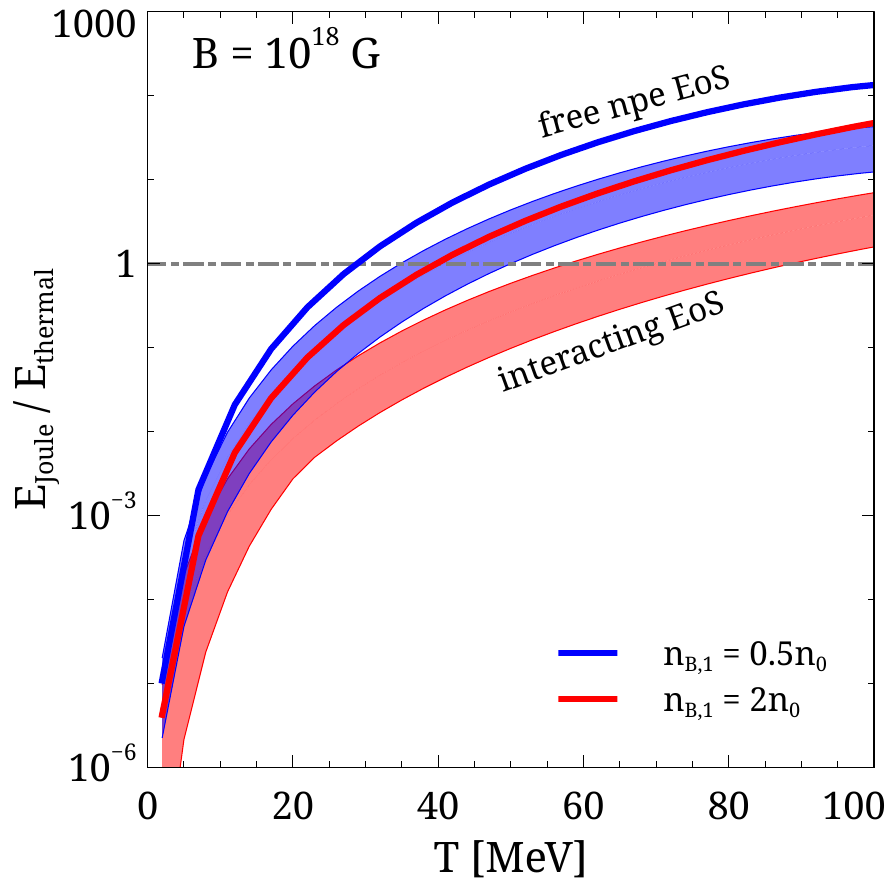}\hspace{1cm}
\includegraphics[width=0.45\textwidth]{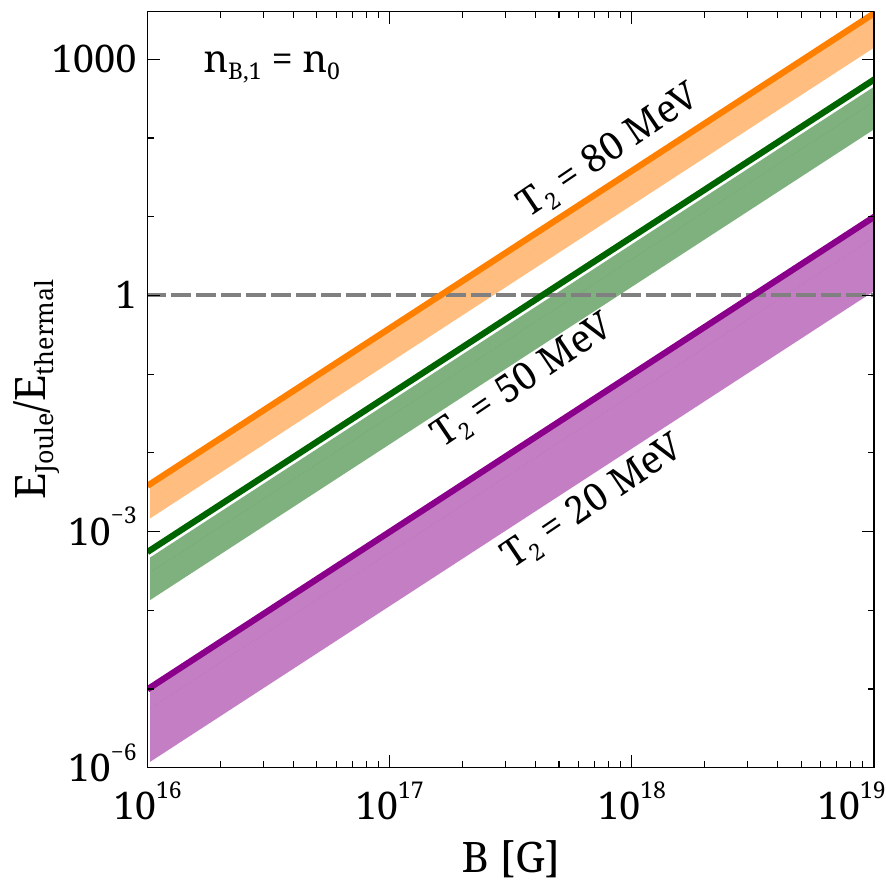}
\caption{Left: The ratio of the Joule heating caused by the chiral effects compared to the increase in thermal energy due to the shock. Right: The ratio of Joule heating to the increase in thermal energy as a function of the magnetic field strength, assuming the upstream matter is at nuclear saturation density.  Three different shock strengths, corresponding to three different downstream temperatures, are shown.  The same conventions apply as in Fig.~\ref{fig:n2_over_n1_interacting}.}
\label{fig:Ejoule}
\end{figure*}

Finally, we calculate the Joule heating using Eqs.~\ref{th2}, \ref{eq:t_weak_interacting}, and \ref{eq:mu5_interacting}
\begin{equation}
    \epsilon_\text{J} = \dfrac{51\pi}{10}\dfrac{G^2\left(1+3g_A^2\right)\mu_n^*\mu_p^*\textbf{B}_0^2\delta\mu^2T^4}{\alpha_{\text{EM}}\log{\alpha_{\text{EM}}^{-1}}m_e^4}.\label{joule_interacting}
\end{equation}
We compare the Joule heating to the thermal energy gained from the shock, which we define as the energy density of the matter at the downstream conditions minus the energy density at the same conditions but with the temperature set to zero (as in Eq.~\ref{eq:shock_heating_definition}).  In the left panel of Fig.~\ref{fig:Ejoule}, we plot the ratio of the Joule heating from chiral effects (Eq.~\ref{joule_interacting}) to the heating that arises from the shock itself.  We are looking for situations where the Joule heating is noticeable compared to the shock without any chirality effects, and this is the case when the shock is strong enough to heat matter, especially low-density matter, to temperatures of at least 30-40 MeV.  However, this assumes a the strongest possible magnetic field set by the QCD (quantum chromodynamics) scale $\sim 10^{18}$ G. In the right panel of Fig.~\ref{fig:Ejoule}, we show the ratio of Joule heating to thermal heating as a function of the magnetic field strength, for matter with an upstream density of $1n_0$.  Three different strength shocks are shown: one that heats matter from zero temperature to 20 MeV, one that is heated to 50 MeV, and one that is heated to 80 MeV.  The Joule heating goes as the square of the magnetic field strength, so the curves are all linear in this log-log plot.  We see that the chiral imbalance generated by strong shocks can lead to substantial Joule heating if the magnetic field is as low as $10^{17}$ G, while weaker shocks require much higher magnetic field strengths.  As numerical simulations frequently see shocks heating localized regions of matter to $T=50 \text{ MeV}$ \cite{Perego:2019adq,Biswas:2026wph,Fields:2023bhs,particles2010004,Raithel:2021hye,Bernuzzi:2024mfx,Hammond:2021vtv,Figura:2021bcn,Most:2021ktk,Gieg:2025ivb,Bernuzzi:2020tgt}, then if the magnetic field is just a couple times $10^{17}$ G, chiral effects would contribute to enhancing the shock heating by at least 10\% and if the magnetic field is larger than several times $10^{17}$ G, the Joule heating could match the shock heating.

Neutron star merger simulations currently predict spatially averaged magnetic field strengths of up to a few times $10^{16}\text{ G}$, with pockets of matter experiencing magnetic fields of up to perhaps a few times $10^{17}\text{ G}$ \cite{Palenzuela:2021gdo,Aguilera-Miret:2023qih,Aguilera-Miret:2025nts,Ciolfi:2019fie,Kiuchi:2023obe,Reboul-Salze:2024jst} (see the review \cite{Ciolfi:2020cpf}).  Magnetic fields, which are quickly amplified in mergers due to the Kelvin-Helmholtz instability \cite{Kiuchi:2015sga} and then perhaps from dynamos \cite{Reboul-Salze:2024jst,Kiuchi:2023obe}, are not fully resolved in merger simulations, which currently have spatial resolutions of, at best, several tens of meters.  The behavior of magnetic fields on the small distance scales like the beta-equilibrating sliver we are considering here is not known.

In summary, we note that all three effects we discuss - a background chiral chemical potential $\mu_5$, the CPI, and Joule heating (Eqs.~\ref{eq:mu5_interacting}, \ref{eq:CPI_t_interacting}, \ref{joule_interacting}) are linearly proportional to the beta equilibration rate $t_{\text{weak}}^{-1}$, which makes clear the significance of weak equilibration in driving these phenomena.  Furthermore, all effects are proportional to some power of $\delta\mu$, so the further out of weak equilibrium the system is driven, the more powerful the effects.

It is interesting to consider which properties of the interacting EoSs have the most impact on the strength of the various chiral effects considered here.  This turns out to be a complicated question, and a full exploration deserves a separate treatment.  Clearly one important factor is the beta-equilibrium value of the proton fraction and how much that changes as a function of density.  At zero temperature, this can be reduced to the symmetry energy.  If the particle content was independent of density and temperature, then the density and temperature increase induced by the shock would not push the system out of beta equilibrium at all and thus there would be no imbalance of left- and right-handed electrons generated.  In constructing the bands displayed in Figs.~\ref{fig:n2_over_n1_interacting}-\ref{fig:Ejoule}, we used about 15 different EoSs from the list compiled by Dutra \textit{et al.} \cite{Dutra:2014qga}, which in terms of the symmetry energy, range from $J\approx 30\text{ MeV}$ for NL$\rho$ \cite{2008ApJ...678.1517F} to $J\approx 37\text{ MeV}$ for NL3 \cite{Lalazissis:1996rd} and in terms of the symmetry energy slope $L$, from QMCRMF4 with $L\approx 31\text{ MeV}$ \cite{Alford:2022bpp} to NL3 with $L\approx 119\text{ MeV}$.

When looking at the EoS bands in our main results (Figs.~\ref{fig:mu5_shock_interacting}-\ref{fig:Ejoule}), which were constructed by taking the largest value and smallest value predicted by the 15 chosen EoSs from \cite{Dutra:2014qga}, for much of the parameter space explored here, the lower bound came from the IUFII EoS \cite{Nandi:2018ami,Harris:2025isp} (similar to IUFSU) or one of the QMCRMF EoSs \cite{Alford:2022bpp}, while the upper bound came mostly from NL$\rho$ \cite{2008ApJ...678.1517F}.  While the symmetry energy may well contribute to this, it is evidently not the whole story.  First of all, quantities like $\mu_5^b$ are proportional to how far the shocked matter is out of beta equilibrium (that is, $\delta\mu$, c.f.~Fig.~\ref{fig:dmu_shock_interacting}), however one must remember that both the density and temperature of the matter increased across the shock and thus both the density-dependence (symmetry energy) and temperature-dependence of the EoS are key factors.  We are not aware of much exploration of the beta-equilibrated particle content as a function of temperature in dense matter.  Secondly, the impact of the EoS on the relationship between upstream and downstream thermodynamic parameters, connected via the RH conditions, is little-studied as far as we know.  We also notice that the strong interactions that modify the EoS from its naive free Fermi gas form tend to decrease all of the effects studied in this paper.  This likely has something to do with the very slowly varying proton fraction of $npe$ matter, but further investigation is needed.

Matter that is subjected to particularly strong shocks is heated to several tens of MeV, at which point the neutrino mean free path grows quite short.  Beta equilibration can begin to take place with neutrinos in the initial state and both Urca processes ($n\leftrightarrow p+e^-+\bar{\nu}_e$ and $e^-+p\leftrightarrow n+\nu_e$ can proceed forward and backward.  In the high-temperature limit where the neutrinos are fully thermalized and form Fermi-Dirac distributions, the beta equilibration rates change from Eq.~\ref{urc} to an expression given in \cite{Alford:2021lpp}.  Generally, the beta equilibration in neutrino-trapped matter is much faster than in neutrino transparent, as the kinematics are less frustrated due to the existence of the neutrino (or antineutrino) Fermi sea.  In neutrino-trapped matter, the definition of beta equilibrium is modified to include the neutrino chemical potential, so the shock takes matter that was in neutrino-transparent beta equilibrium and compresses and heats it and the neutrino-trapped Urca processes must adjust the particle content to the neutrino-trapped beta equilibrium condition.  This ``target'' depends on the density of neutrinos that happened to be accumulated in that part of the matter since the time that their mean free path shortened enough to prevent their escape.  We have begun calculations of the key quantities in this paper - background $\mu_5$, CPI growth timescale, and Joule heating - and our preliminary findings are that the results are quite similar to the neutrino-transparent results shown in this paper, although the neutrino-trapped results depend strongly on the neutrino density assumed.  Finally, at intermediate temperatures (perhaps for temperatures less than about 30 MeV), neutrinos play a role in beta equilibration, but are not in thermal equilibrium.  Neutrino flux from other regions of the merger also must be taken into account.  Rates in this intermediate regime are much more difficult to calculate and a full neutrino-transport scheme in the context of a neutron star merger simulation is required \cite{Alford:2026kwd}.

At zero temperature, muons appear in the ground state of dense matter just about nuclear saturation density.  When they appear, they reduce the electron density, as some electrons convert into muons, but increase the proton fraction \cite{Loffredo:2022prq}.  Muons are charged fermions, and thus like electrons they could develop a chiral imbalance, but this is unlikely to be significant because the muon mass is much larger than the electron mass and the chiral flip rate goes as the square of the mass (Eq.~\ref{ndg} or \ref{dg}).  Likely, the influence of muons would be in their impact on the EoS and on the electron density and therefore the corresponding beta equilibration rates.  The results obtained in this paper are sensitive to the beta equilibration rates and thus the inclusion of muons should be investigated in the future.

\section{Conclusion} 
In this paper our goal was to explore  strong and abrupt density fluctuation in dense medium and its impact on chiral transport phenomena fueled by the chiral magnetic effect. The setting of shockwaves in dense matter allows us to address these effects concretely. We construct for the first time, a theoretical framework which combines the physics of weak interaction, the chiral magnetic effect, and the physics of shock waves.  In the physical picture that we provide, a shock wave compresses a downstream region, pushing it out of weak equilibrium.  As the Urca processes restore weak equilibrium, a chiral imbalance is generated, leading possibly to CME effects.  Our estimates show that for strong enough shocks, i.e.~if the shock heating and compression are large enough, the resulting weak interactions can give rise to a significant chiral imbalance, overwhelming chirality flipping due to the finite electron mass. Furthermore, the process can sustain a chiral imbalance for sufficient time to either fuel a chiral plasma instability or heat up the medium through significant ohmic dissipation in the presence of a strong background magnetic field.  These are small-scale effects in neutron star merger conditions.  The thickness of the sliver where the effects happen is sub-meter, which is well below the tens-of-meters resolution of current simulations \cite{Kiuchi:2023obe,Radice:2024gic,Zappa:2022rpd}.  At lower temperatures, the beta-equilibration rate slows and therefore the sliver grows to tens, hundreds, or thousands of meters, but no substantial chiral imbalance exists there because the beta equilibration is so slow.  

We illustrated the associated concepts and estimates using non-interacting EoS and restricting to two specific conditions of upstream parameters. One of the important lessons of the exercise was that shock jumps that drive the system further out of weak equilibrium while generating higher temperatures are able to sustain the CPI instability as well as produce strong Joule heating when the medium is already highly magnetized.
These results were complemented using interacting EoSs to capture more realistic NS environment. We find that the effect of including interactions is to produce less dramatic departure from weak equilibrium, which reduces the parameter range over which CPI survives or Joule heating is substantial. However, both effects still survive in realistic range of parameters found in NS mergers.  It's also important to note that 
the nuclear EoS is well constrained around $n_0$ at low temperature, but at higher densities and temperatures, as well as far out of weak equilibrium, the uncertainty grows dramatically. The chiral effects on the other hand get realized in downstream regions at relatively high temperature, and dramatically out of equilibrium, precisely where the EoSs are less reliable. Thus, the lesson from our analysis is that chiral effects can be indeed be substantial. However, to pinpoint the exact range of parameters where this is the case one requires better constraints on the EoS at high density, high temperature and out of weak equilibrium.  Within the limitations imposed by these uncertainties, our results offer an evaluation of the role of chiral effects in the context of existing equations of state.

Finally, it is important to emphasize that the chiral effects we have identified are sourced by departures from weak equilibrium. Although we generated such conditions through shock waves, this mechanism is not essential. Any violent event involving density or thermal fluctuations that drives the system out of weak equilibrium can, in principle, produce the same chiral effects. It would be interesting to explore in future work whether alternative mechanisms can similarly generate such departures and thereby induce comparable chiral phenomena. 

Another interesting direction of future work will be to explore the shockwave dynamics and associated CME effects described by Fig.~\ref{shock-pic} in numerical simulations. In fact, it is worthwhile exploring this dynamics while varying the electron mass considering it as a parameter as opposed to fixed at its physical value. In other words, it is interesting to assess whether numerical simulations can capture the physics describe in Fig.~\ref{shock-pic} for physical values of parameters and beyond.

Several improvements are needed to make more definitive conclusions about realistic environments. Firstly, improved temperature constraints on the EoSs including interactions would help pinning down the parameter space where chiral effects are important.  In particular, new particle degrees of freedom are likely to appear at high temperatures and densities (for example, muons \cite{Loffredo:2022prq,Pajkos:2024iry,Gieg:2024jxs}, pions \cite{Fore:2019wib,Vijayan:2023qrt,Pajkos:2024iry,Harris:2024ssp} and hyperons \cite{Kochankovski:2025lqc,Alford:2020pld}), and it is also possible that a substantial density jump may drive matter into a regime that is described by deconfined quark matter \cite{Verma:2022pcg,PhysRevC.84.065805,PhysRevC.91.055806}. Thus, it is also worth exploring how a possible phase transition across a shock front may affect our estimates.  

Neutrinos and sophistication related to the Urca rates represent probably the biggest ``known'' physics that can be improved or added to our analysis.  The Urca rates should, ultimately, be calculated by doing a full integration over the phase space \cite{Alford:2018lhf,Alford:2021ogv} instead of the degenerate limit expression used here (Eq.~\ref{urc}).  This is likely to reduce temperature-dependence of the Urca rate as matter becomes nondegenerate at temperatures of several tens of MeV. However the beta equilibration rate also depends on $\delta\mu$ (c.f.~the degenerate expression Eq.~\ref{urc}) and it is unclear without a detailed calulcation how the ultimate equilibration rate would change if the $\delta\mu$ term were to dominate. In addition, the effects of strong magnetic fields should be included in the Urca rate calculations, though this only effects the rate at low temperatures \cite{Tambe:2024usx,Tambe:2025evw,Kumamoto:2024jiq}.  Next, the shock heating quickly pushes the matter into a density and temperature regime where the neutrino mean free path is quite short \cite{Alford:2018lhf,Roberts:2016mwj}.  On some timescale \cite{Espino:2023dei}, neutrinos therefore become thermally and chemically equilibrated in the shocked matter, and thus change the matter composition and the beta equilibration rate, and this should be incorporated into the analysis presented here.  

Finally, our analysis applies to shock fronts for which shock thickness is much smaller than the out of equilibrium region over which weak interaction takes place.   This allows us to keep particle content frozen across the shock. Our analysis should be extended to cases where the two scales become comparable and beta equilibration may need to be incorporated in the calculation of the Rankine-Hugoniot conditions \cite{landau1987fluid}. Many shocks in supernovae and neutron star mergers occur at very low densities \cite{Muller:2020ard}, well below neutron drip, where the Urca rate that we use in clearly inapplicable, and therefore the production of chiral imbalances in this low-density matter deserves a separate study.  
\section*{Acknowledgments}
We would like to thank David Kaplan, David Radice, Sanjay Reddy, Varun Vaidya, Naoki Yamamoto and Laurence Yaffe for insightful discussions.  S.P.H.~acknowledges the support of the National Science Foundation grants PHY-2116686 and PHY-2621752.  S.S.~acknowledges support from the U.S.~Department of Energy, Nuclear Physics Quantum Horizons program through the Early Career Award DE-SC0021892.  
\appendix
\section{Finite-temperature corrections to chiral charge density relationship}\label{appendix}
\begin{figure}\centering
\includegraphics[width=0.45\textwidth]{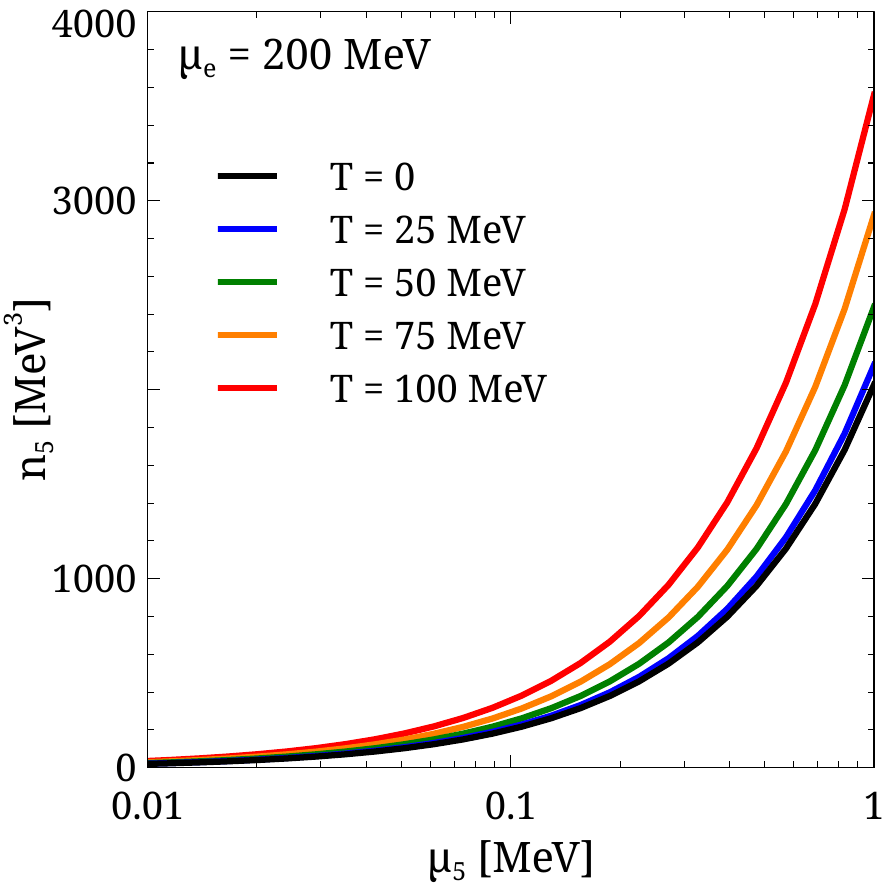}
\caption{Zero and finite temperature expressions for $n_5$ as a function of $\mu_5$.  }
\label{fig:n5_vs_mu5_finiteT}
\end{figure}
At zero temperature, we can see from Eq.~\ref{eq:n_left_right_zerotemp} that 
\begin{equation}
    n_5 = \dfrac{\mu_5}{2\pi^2}\left(\mu_e^2+\dfrac{\mu_5^2}{12}\right) \quad\text{(zero temperature)}.
\end{equation}
At finite temperature, 
\begin{align}
    n_5 &= \dfrac{1}{2\pi^2}\int_0^{\infty}\mathop{dk}k^2\left(\dfrac{1}{1+e^{\beta\left(k-\mu_{e,R}\right)}}-\dfrac{1}{1+e^{\beta\left(k-\mu_{e,L}\right)}}\right)\nonumber\\
    &=\dfrac{T^3}{\pi^2}\left\{ \phi_3\left[ -e^{\beta\left( \mu_e-\mu_5/2\right)} \right] -\phi_3\left[ -e^{\beta\left( \mu_e+\mu_5/2\right)} \right]    \right\}. 
\end{align}
For a typical choice of $\mu_e$ (200 MeV), in Fig.~\ref{fig:n5_vs_mu5_finiteT} we compare the zero-temperature and finite-temperature expressions for $n_5$ as a function of $\mu_5$.  We only consider $\mu_5$ below 1 MeV, because in the mechanism described in this paper, the shocked matter will evolve from $\mu_5=0$ to the equilibrium (background) value of $\mu_5^b$, which according to Fig.~\ref{fig:mu5_shock_interacting} is usually less than 1 MeV.  The finite-temperature expression, for temperatures up to 100 MeV, is larger than the zero-temperature expression, but by less than a factor of 2.  For $\mu_5$ small compared to $\mu_e$, the finite-temperature expression for $n_5$ is larger than the zero-temperature expression by a factor of $-2\left(T/\mu_e\right)^2\phi_2\left(-e^{\mu_e/T}\right)$.

\bibliography{ref.bib}
\end{document}